%% file: SMP-22-010_temp.tex
\pdfoutput=1
\documentclass[11pt,twoside,a4paper,cmspaper,final,collab]{cms-tdr}
\def\svnVersion{55c3579}\def\svnDate{2025/05/08}\def\cmsCernNoTag{CERN-EP-2024-208}\def\cmsCernDate{\today}\def\cmsMessage{Published in Physics Letters B as \href{http://dx.doi.org/10.1016/j.physletb.2025.139526}{\doi{10.1016/j.physletb.2025.139526}.}}
\begin{document}\cmsNoteHeader{SMP-22-010}

\newlength\cmsFigWidth
\ifthenelse{\boolean{cms@external}}{\setlength\cmsFigWidth{\columnwidth}}{\setlength\cmsFigWidth{0.497\textwidth}}

\newlength\cmsFigWide
\ifthenelse{\boolean{cms@external}}{\setlength\cmsFigWide{\columnwidth}}{\setlength\cmsFigWide{0.65\textwidth}}

\newlength\cmsTabSkip\setlength\cmsTabSkip{1.5ex}

\ifthenelse{\boolean{cms@external}}{\providecommand{\cmsLeft}{upper\xspace}}{\providecommand{\cmsLeft}{left\xspace}}
\ifthenelse{\boolean{cms@external}}{\providecommand{\cmsRight}{lower\xspace}}{\providecommand{\cmsRight}{right\xspace}}
\ifthenelse{\boolean{cms@external}}{\newcommand{\cmsTable}[1]{\resizebox{\columnwidth}{!}{#1}}}{\newcommand{\cmsTable}[1]{#1}}

\newcommand{\sinw} {\ensuremath{\sin^2\theta_\mathrm{W}}\xspace}
\newcommand{\sineff} {\ensuremath{\sin^2\theta_\text{eff}^{\Pell}}\xspace}
\newcommand{\afour} {\ensuremath{A_4}\xspace}
\newcommand{\afourym} {\ensuremath{A_4(\abs{Y}, M)}\xspace}
\newcommand{\cs} {\ensuremath{\cos\theta_\mathrm{CS}}\xspace}
\newcommand{\phiCS} {\ensuremath{\phi_\mathrm{CS}}\xspace}
\newcommand{\tautau} {\ensuremath{{\tau\tau}}\xspace}
\newcommand{\afb} {\ensuremath{A_\mathrm{FB}}\xspace}
\newcommand{\afbw} {\ensuremath{A_\mathrm{FB}^\mathrm{w}}\xspace}
\newcommand{\afbwym} {\ensuremath{A_\mathrm{FB}^\mathrm{w}(\abs{y},m)}\xspace}
\newcommand{\mt} {\ensuremath{m_\mathrm{T}}\xspace}
\newcommand{\feta} {\ensuremath{\abs{\eta}}\xspace}
\newcommand{\uu} {\ensuremath{\PGm\PGm}\xspace}
\newcommand{\ue} {\ensuremath{\PGm\Pe}\xspace}
\newcommand{\eu} {\ensuremath{\Pe\PGm}\xspace}
\newcommand{\ee} {\ensuremath{\Pe\Pe}\xspace}
\newcommand{\myeg} {\ensuremath{\Pe\Pg}\xspace}
\newcommand{\eh} {\ensuremath{\Pe\Ph}\xspace}
\newcommand{\ug} {\ensuremath{\PGm\Pg}\xspace}
\newcommand{\uh} {\ensuremath{\PGm\Ph}\xspace}
\newcommand{\zll} {\ensuremath{\PZ/\PGg\to\Pell\Pell}\xspace}
\newcommand{\wjet} {\ensuremath{\PW\!+\!\text{jets}}\xspace}
\newcommand{\ztt} {\ensuremath{\PZ/\gamma\to\tautau}\xspace}
\newcommand{\minnlo}{\textsc{MiNNLO$_\mathrm{PS}$}\xspace}
\newcommand{\powhegzew}{\POWHEG-\textsc{z}\_ew\xspace}

\newcommand{\pdf}{\ensuremath{\,\text{(PDF)}}\xspace}
\newcommand{\exper}{\ensuremath{\,\text{(exp)}}\xspace} 
\newcommand{\xfitter}{\ensuremath{\textsc{xFitter}}\xspace}

\cmsNoteHeader{SMP-22-010}

\title{Measurement of the Drell--Yan forward-backward asymmetry and of the effective leptonic weak mixing angle in proton-proton collisions at \texorpdfstring{$\sqrt{s} = 13\TeV$}{sqrt{s} = 13 TeV}}

\date{\today}

\abstract{The forward-backward asymmetry in Drell--Yan production and the effective leptonic electroweak mixing angle are measured in proton-proton collisions at $\sqrt{s} = 13$\TeV, collected by the CMS experiment and corresponding to an integrated luminosity of 138\fbinv. The measurement uses both dimuon and dielectron events, and is performed as a function of the dilepton mass and rapidity. The unfolded angular coefficient \afour is also extracted, as a function of the dilepton mass and rapidity. Using the CT18Z set of parton distribution functions, we obtain $\sineff = 0.23152 \pm 0.00031$, where the uncertainty includes the experimental and theoretical contributions. The measured value agrees with the standard model fit result to global experimental data. This is the most precise \sineff measurement at a hadron collider, with a precision comparable to the results obtained at LEP and SLD.}

\hypersetup{%
pdfauthor={CMS Collaboration},%
pdftitle={Measurement of the Drell--Yan forward-backward asymmetry and of the effective leptonic weak mixing angle in proton-proton collisions at sqrt{s} = 13 TeV},%
pdfsubject={CMS},%
pdfkeywords={CMS, electroweak, mixing angle, Drell-Yan}
}

\maketitle

\section{Introduction}

The electroweak (EW) mixing angle,
relating the masses of the \PW and \PZ bosons ($m_{\PW}$ and $m_{\PZ}$),
$\sinw = 1 - m_{\PW}^2/m_{\PZ}^2$,
is a key parameter in the standard model (SM) of particle physics.
It defines the ratio between the vector and axial-vector couplings of the \PZ boson with fermions,
$v_\mathrm{f} / a_\mathrm{f} = 1 - 4 \abs{Q_\mathrm{f}} \sinw$,
where $Q_\mathrm{f}$ is the electric charge of the fermion~f.
Although EW radiative corrections change this tree-level picture,
the same relation can be kept for all orders by introducing
an effective mixing angle that depends on the fermion flavor.
Both $m_{\PW}$ and the effective leptonic mixing angle (\sineff)
can be precisely obtained in the SM, 
using LEP and SLD measurements at the \PZ resonance~\cite{lepsld}, 
the muon lifetime, and the masses of the top quark and of the Higgs boson.
The current predictions, from a global EW fit~\cite{Workman:2022ynf}, are 
$m_{\PW} = 80\,360 \pm 6$\MeV and $\sineff = 0.23155 \pm 0.00004$.
Measurements of $m_{\PW}$ and \sineff provide an important test of the SM since
deviations from the predictions could be evidence of beyond-the-SM physics.
Currently, the most precise \sineff values are $0.23221 \pm 0.00029$, 
from \PQb quark forward-backward asymmetry ($A_\mathrm{FB}^{0,\PQb}$) results at the CERN LEP experiments, 
and $0.23098\pm0.00026$, 
from left-right asymmetry ($A_\mathrm{LR}^{0}$) data at the SLD experiment at SLAC~\cite{lepsld}.
These two values differ by 3.2 standard deviations.
This Letter presents a \sineff measurement with a precision that exceeds 
that of all previous hadron collider measurements~\cite{Abazov:2008xq,
Abazov:2011ws,
Chatrchyan:2011ya,
Aaltonen:2013wcp,
Aaltonen:2014loa,
Abazov:2014jti,
Aad:2015uau,
Aaij:2015lka,
Aaltonen:2016nuy,
Abazov:2017gpw,
Aaltonen:2018dxj,
cms:wma}
and is comparable to that of the LEP and SLD results~\cite{lepsld}.

In proton-proton ($\Pp\Pp$) collisions at the CERN LHC, 
Drell--Yan (DY) production is driven by the vector and axial-vector couplings of the \PZ boson to fermions.
We measure \sineff using 
$\Pp\Pp \rightarrow Z/\gamma +X \rightarrow \Pell^+\Pell^- + X$ events, 
where the $\Pell^+\Pell^-$ dilepton invariant mass is near the \PZ boson mass and
$\Pell$ is either a muon or an electron.
To reduce theoretical and experimental uncertainties, we use leptonic angular distributions
in the Collins--Soper~(CS) frame~\cite{Collins:1977iv}.
The angle of the negatively charged lepton relative to the incoming quark direction,
$\theta_\mathrm{CS}$, is computed as 
\begin{equation*}
\cs = \frac{2 \, (Q_1^+Q_2^- - Q_1^-Q_2^+)}{\sqrt{m^2 (m^2 + p^2_\mathrm{T})}} \frac{y}{\abs{y}},
\end{equation*}
where $m$, \pt, and $y$ are the 
mass, transverse momentum, and rapidity of the dilepton, respectively, 
and $Q_i^\pm = (E_i \pm P_{z,i}) / \sqrt{2}$,
with $E$ and $P_z$ being the energy and longitudinal momentum
of the leptons, with $i=1$ for $\Pell^-$ and 2 for $\Pell^+$.
In this equation, $y$ is used as a proxy for the valence quark direction.
Before final-state radiation (FSR), the DY differential cross section
for the polar ($\theta$) and azimuthal ($\phi$) decay angles  
can be described~\cite{PPHEP} as
\begin{equation*}
\frac{16\pi}{3\,\sigma} \, \frac{\mathrm{d}\sigma}{\mathrm{d}\cos\theta\,\mathrm{d}\phi} =
1+\cos^2\theta + \sum_{i=0}^{7} A_i \, f_i(\theta,\phi) ,
\end{equation*}
where $\sigma$ is the unpolarized cross section and
$f_0 = 0.5 (1-3\cos^2\theta)$, 
$f_1 = \sin2\theta\cos\phi$,
$f_2 = 0.5\sin^2\theta\cos2\phi$,
$f_3 = \sin\theta\cos\phi$,
$f_4 = \cos\theta$,
$f_5 = \sin^2\theta\sin2\phi$,
$f_6 = \sin2\theta\sin\phi$, and
$f_7 = \sin\theta\sin\phi$.
The angular coefficients $A_i$ are functions of the
mass ($M$), transverse momentum ($P_\mathrm{T}$), and rapidity ($Y$) of the pre-FSR dilepton.
Except for the forward-backward asymmetry coefficient, \afour, all the angular coefficients are small.
Integrating the cross section over $\phi$, the
dependence on $\cos\theta$ is fully described by the $A_0$ and \afour coefficients.
In this analysis, we also integrate over $P_\mathrm{T}$.

The \sineff analysis of $\sqrt{s} = 8$\TeV CMS data~\cite{cms:wma} used the 
directly measured forward-backward angular-weighted asymmetry, \afbw,
approximately equal to
$\afb = (\sigma_\mathrm{F} - \sigma_\mathrm{B}) / (\sigma_\mathrm{F} + \sigma_\mathrm{B})$ 
in full phase space,
where $\sigma_\mathrm{F}$ and $\sigma_\mathrm{B}$ 
are the $\cs > 0$ and $\cs < 0$ integrated cross sections, respectively.
In full phase-space and before FSR, $\afb = (3/8) \, \afour$.
This angular-weighted asymmetry method~\cite{Bodek:2010qg},
with weights that are functions of the rapidity and \cs of the dilepton,
uses reconstructed variables
and benefits from the cancelation of systematic uncertainties in
the detection acceptance and efficiencies, 
which are approximately identical for positive and negative \cs. 
In this Letter, we report a \sineff measurement with reduced uncertainties,
obtained by applying this method to the larger $\sqrt{s} = 13$\TeV CMS data sample.
We also report \sineff results obtained by fitting the unfolded angular coefficient \afour
as a function of the rapidity and mass of the pre-FSR dilepton.
This second method, while more complex, 
provides a measurement with smaller theory and 
parton distribution function (PDF) uncertainties, 
which can be used in reinterpretations of the results (\eg, with improved PDFs) 
and in combinations with other measurements.

\section{The CMS detector and data samples}

The CMS apparatus~\cite{CMS:2008xjf,CMS:2023gfb} is a multipurpose and nearly hermetic detector, 
designed to trigger on~\cite{CMS:2020cmk,CMS:2016ngn} and identify electrons, muons, photons, 
and (charged and neutral) hadrons~\cite{CMS:2020uim,CMS:2018rym,CMS:2014pgm}. 
A global event description algorithm~\cite{CMS:2017yfk} aims to reconstruct all individual particles in an event, 
combining information provided by the all-silicon inner tracker and by 
the crystal electromagnetic~(ECAL) and brass-scintillator hadron~(HCAL) calorimeters, 
operating inside a 3.8\unit{T} superconducting solenoid, 
with data from the gas-ionization muon detectors embedded in the flux-return yoke outside the solenoid. 

The ECAL and HCAL calorimeters provide the pseudorapidity coverage of $\abs{\eta} < 3.0$,
which is extended up to $\abs{\eta} = 5$ by forward calorimeters that use quartz fibers as active material~\cite{Bayatian:2006jz};
electromagnetic and hadronic showers can be distinguished by comparing the signals 
of long fibers that extend over the full depth of the detector with those of short fibers 
that start 22\cm into the detector.
The reconstructed particles are used to build \PGt leptons, jets, 
and missing transverse momentum~\cite{CMS:2018jrd,CMS:2016lmd,CMS:2019ctu}.

The measurement reported in this Letter is based on
samples of $\Pp\Pp$ collisions at $\sqrt{s} = 13\TeV$,
collected by the CMS experiment in 2016--2018.
The 2016 data sample is split in two, referred to as 2016a and 2016b,
because of a change made to improve the hit efficiency of the silicon tracker 
at high instantaneous luminosities~\cite{CMS:2021ime}.
The integrated luminosities of the 2016a, 2016b, 2017, and 2018 data samples 
are 19.5, 16.8, 41.5, and 59.8\fbinv, respectively,
with 1.2--2.5\% individual uncertainties~\cite{CMS-LUM-17-003, CMS-PAS-LUM-17-004, CMS-PAS-LUM-18-002}.

The events are selected by dilepton and single-lepton triggers.
Both triggers include a lepton isolation requirement.
The dimuon trigger requires one muon with $\pt > 17$\GeV and another with $\pt > 8$\GeV,
whereas the corresponding thresholds for the dielectron trigger are 23 and 12\GeV.
The single-muon and single-electron triggers
impose minimum \pt thresholds of 24 and 27\GeV (28\GeV in 2018), respectively.
During part of the data-taking periods, these single-lepton triggers were prescaled,
with the unprescaled triggers having higher \pt thresholds.
The muons are reconstructed within the $\abs{\eta} < 2.4$ range.

Three electron categories are considered, 
depending on the detectors involved in their reconstruction: 
central electrons (\Pe), in the $\abs{\eta}$ range covered by the silicon tracker;
forward ECAL electrons (\Pg), 
within $2.5 < \abs{\eta} < 2.87$;
and forward HCAL electrons (\Ph) 
in the 3.14--4.36 $\abs{\eta}$ range.
We consider four dilepton categories: \uu, \ee, \myeg, and \eh.
The central-central \uu and \ee channels use both single-lepton and dilepton triggers, 
whereas the \myeg and \eh central-forward channels only use single-electron triggers.
In events with two or more central leptons satisfying the selection criteria,
the central-central dileptons are composed of the two highest \pt leptons (``leading" and ``trailing").
The two leptons are required to satisfy the kinematic requirements listed in Table~\ref{tab_acceptance},
have opposite charges, and satisfy trigger-matching criteria: 
either they are both matched with dilepton trigger objects
or one of them (with $\pt > 25$\GeV for muons and 30\GeV for electrons)
is matched with a single-lepton trigger object.
For the \eh events, the central electron is required to be in the endcap~$(\abs{\eta} >1.57)$
to reduce background contamination.

\begin{table}[t] 
\centering
\topcaption{The lepton $\eta$ and \pt acceptance windows 
applied in the four measurement channels.
The 1.44--1.57 \feta range, between the barrel and endcap ECAL, is excluded for central electrons.}
\label{tab_acceptance}
\begin{tabular}{ l  c  c  c  c  }
 & \multicolumn{2}{c}{\feta}   & $p_\text{T,min}^\text{leading}$  &  $p_\text{T,min}^\text{trailing}$  
\\ \hline
\uu   & \multicolumn{2}{c}{0.00--2.40}  &  20\GeV &  10\GeV \\
\ee   & \multicolumn{2}{c}{0.00--2.50}  &  25\GeV &  15\GeV \\[2mm]
 & $\abs{\eta_\Pe}$  &  $\abs{\eta_{\Pg,\Ph}}$   & $p_\text{T,min}^{\Pe}$    &  $p_\text{T,min}^{\Pg,\Ph}$  \\ \hline
\myeg   & 0.00--2.50   &  2.50--2.87  &  30\GeV &  20\GeV \\
\eh   & 1.57--2.50   &  3.14--4.36  &  30\GeV &  20\GeV \\
\end{tabular}
\end{table}

The reconstructed muons are required to pass the ``standard medium-identification" 
and ``loose tracker isolation" criteria, with an efficiency of about 95\% ~\cite{CMS:2018rym,CMS:2021yvr}.
The central electron candidates must satisfy the cut-based ``medium-identification criteria",
including an isolation requirement, with an efficiency of about 80\%~\cite{CMS:2020uim}.
In the \myeg and \eh event classes, 
the central electron must also satisfy the ``selective charge identification" criterion, which has about 95\% efficiency 
and requires that three different charge determination methods give the same result (in \ee events, this is not required, 
and each electron charge is determined by a ``majority'' rule~\cite{CMS:2020uim}).

The \Pg and \Ph electrons are reconstructed and selected using independent procedures,
to optimise their individual signal-selection efficiencies.
Dedicated multivariate identification criteria, based on the shower shape and isolation variables, were developed for the \Pg electrons.
The cut-based working point with a signal selection efficiency of about 80\% was derived using TMVA~\cite{tmva}.
The \Ph electrons are reconstructed from the constituents of jets 
built with the anti-\kt algorithm~\cite{Cacciari:2008gp}, 
using a cone radius of 0.4.
Several reconstruction and identification algorithms were developed 
to optimally reconstruct the energy and direction of the prompt \Ph electrons, 
and to distinguish them from the quark and gluon jets. 
In particular, we use multivariate classification and regression
models implemented in \textsc{TensorFlow}~\cite{tensorflow2015-whitepaper},
reaching better resolutions than with simpler methods.
The input variables include, in addition to some pileup-sensitive variables, 
all the individual long and short fiber read-out channels of each $\eta$-$\phi$ cell in a jet. 
Therefore, they contain isolation and shower shape (both transverse and longitudinal) information. 
The working point corresponds to an electron selection efficiency of about 70\%.
In the \myeg and \eh classes,
the two electrons must be in the same pseudorapidity hemisphere;
the charge of the forward electron cannot be measured and is assumed to be 
the opposite of the charge of the central electron.

After applying the selection criteria described above, we retain about 
111 million \uu, 59 million \ee,  5.0 million \myeg, and 3.3 million \eh events, 
including small background contaminations, as discussed in Section~\ref{sec_syst}.
The number of \ee event is smaller than that of \uu given the lower electron selection efficiencies
and higher electron \pt cuts, driven by the online trigger thresholds.

\begin{figure}[tbh]
    \centering
    \includegraphics[width=\cmsFigWidth]{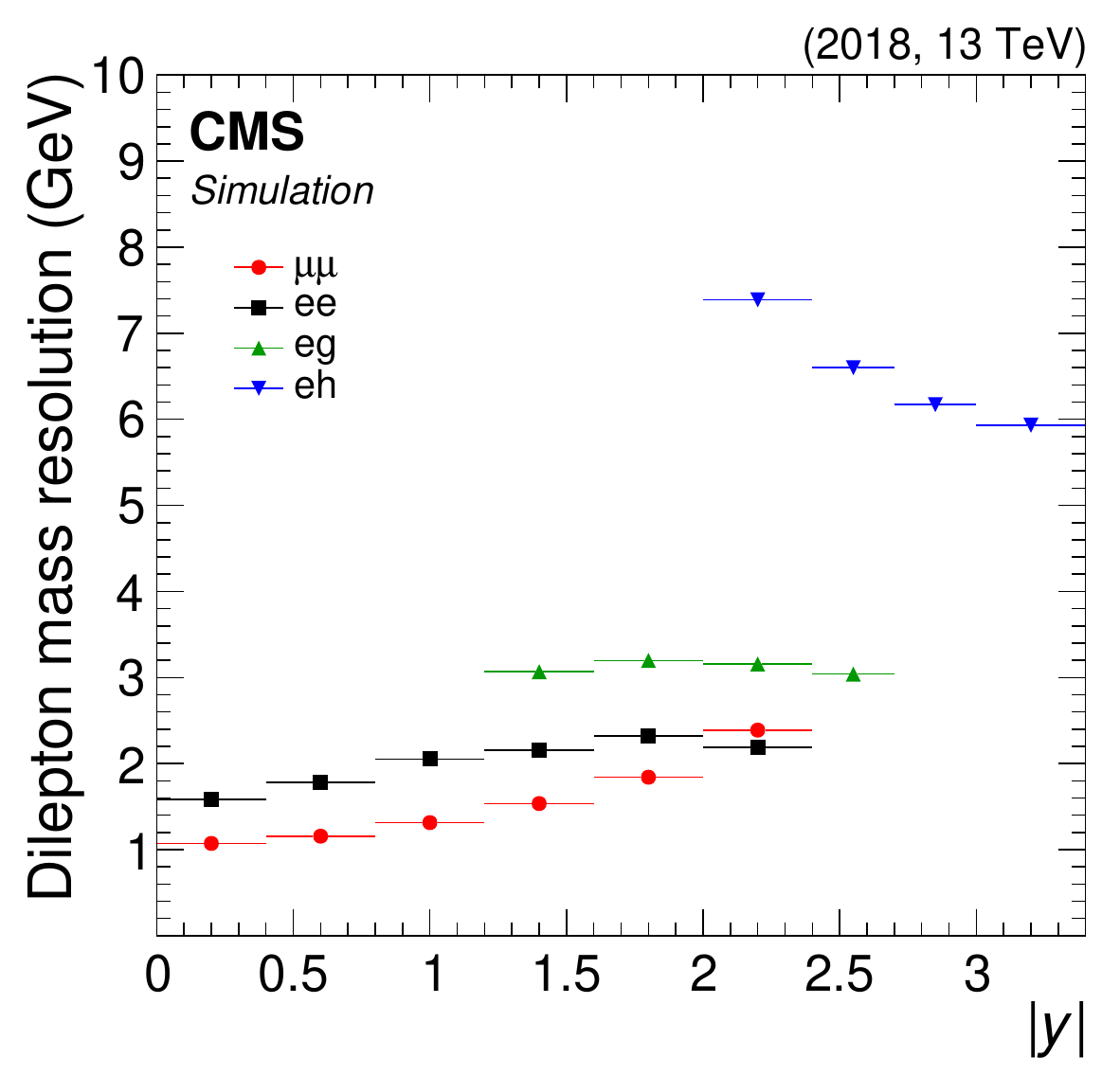}
    \caption{Dilepton mass resolution, at the \PZ mass, as a function of the dilepton absolute rapidity,
    for the \uu, \ee, \myeg, and \eh channels.}
    \label{fig:resolution-vs-rapidity}
\end{figure}
Figure~\ref{fig:resolution-vs-rapidity} shows how the dilepton mass resolution, at the \PZ mass,
varies with absolute rapidity, for the four dilepton categories.

\section{Simulation and corrections}

The signal and background processes are modeled using Monte Carlo (MC) simulation. 
The \zll signal ($\Pell=\Pe,\PGm$) and background ($\Pell=\PGt$) events 
are generated with the \POWHEG2.0 event generator~\cite{Frixione:2007vw, Alioli:2010xd},
with a multi-scale improved next-to-next-to-leading order (\minnlo)
accuracy in quantum chromodynamics 
(QCD)~\cite{Monni:2019whf,Monni:2020nks}, 
matched with \PYTHIA8.2~\cite{Sjostrand:2014zea} 
for the parton shower and hadronization steps, 
as well as for initial-state radiation. 
Photon FSR is simulated with 
the \PHOTOS2 package~\cite{Barberio:1993qi, Golonka_2006}.
The signal events are generated with the 
next-to-next-to-leading order (NNLO) NNPDF3.1 PDF set~\cite{NNPDF:2017mvq}. 
Alternative event weights for variations of renormalization and factorization scales, 
PDF sets~\cite{NNPDF:2021njg, Hou:2019efy, Bailey:2020ooq}, 
and their uncertainties are also considered~\cite{Butterworth:2015oua}.
Additional small corrections to \minnlo are described below.

Top quark pairs, $s$-channel single top events, and \wjet events 
are generated with the event generator 
\MGvATNLO~2.6.5~\cite{Alwall:2014hca},
at next-to-leading order (NLO) accuracy in QCD. 
The $t$-channel single top and $\PQt$\PW events are generated 
with \POWHEG, at NLO in QCD.
The diboson event categories $\PZ\PZ$, $\PW\PZ$, $\PW\PW$, and $\gamma\gamma \to \Pell\Pell$ events 
are generated with \PYTHIA8 at leading order (LO).
In all these cases, \PYTHIA8 is used to model the parton shower and hadronization steps, 
as well as initial and final state radiation.
As discussed in Section~\ref{sec_backgrounds}, 
some of the backgrounds are evaluated using data driven methods and do not require MC samples. 

The detector response is emulated with \GEANTfour~\cite{AGOSTINELLI2003250}.
All the simulated final-state particles are reconstructed with the same algorithms as used for the measured data
and the dilepton event candidates are selected with the criteria described in the previous section.
Additional $\Pp\Pp$ interactions in the same and nearby bunch crossings, ``pileup", 
are included in the simulated samples,
with weights based on the measured instantaneous luminosity and total
inelastic cross section~\cite{Sirunyan:2303548}.
Various other corrections are applied to the simulated events.
The signal $\zll$ events are generated at LO EW accuracy, 
with \sineff, $m_\PZ$, and the Fermi constant $G_\mu$ as input parameters,
and are corrected to account for NLO virtual weak corrections,
including certain universal higher-order (HO) contributions,
using \powhegzew~\cite{Barze:2013fru,Chiesa:2019nqb,Chiesa:2024qzd}. 
A large event sample is generated at NLO in QCD and LO in EW
with the same EW input scheme as \minnlo and including weights for the NLO+HO EW corrections.
The dilepton $\abs{y}$-$m$-\pt triple-differential cross sections and angular coefficients, 
calculated at LO and NLO+HO EW, are used to correct the corresponding \minnlo predictions.

During the 2016 and 2017 data-taking periods,
a gradual shift in the timing of the inputs of the ECAL L1 trigger in the $\abs{\eta} > 2.0$ region
caused a trigger inefficiency, denoted as ``prefiring''~\cite{CMS:2021yvr,CMS:2020cmk}.
The trigger missed between 10\% and 20\% of the events containing
an electron (of $\pt \gtrsim 50$\GeV) or a jet (of $\pt \gtrsim 100$\GeV) in the $2.5 < \abs{\eta} < 3.0$ range,
the exact loss depending on the \pt and $\eta$ of the trigger object.
Correction factors were computed from data and applied to the acceptance evaluated by simulation.

The lepton selection efficiencies, in the measured and simulated events, 
are evaluated with the tag-and-probe method using \zll events~\cite{CMS:2011aa},
separately for the reconstruction, identification, and trigger selection steps.
In the case of the trigger, they are also evaluated separately for the 
prescaled and unprescaled single-lepton triggers, and for the two dilepton trigger legs. 
The event selection efficiency is computed from these individual components in \pt and $\eta$ bins,
accounting for the correlations between the single-lepton and dilepton trigger selection efficiencies, 
and also including a pileup-dependent factor for the prescaled single-lepton triggers. 
The ratio between the selection efficiencies in the measured and simulated samples
is then applied as a weight to the simulated events.

\begin{figure}[tbh]
    \centering
    \includegraphics[width=\cmsFigWidth]{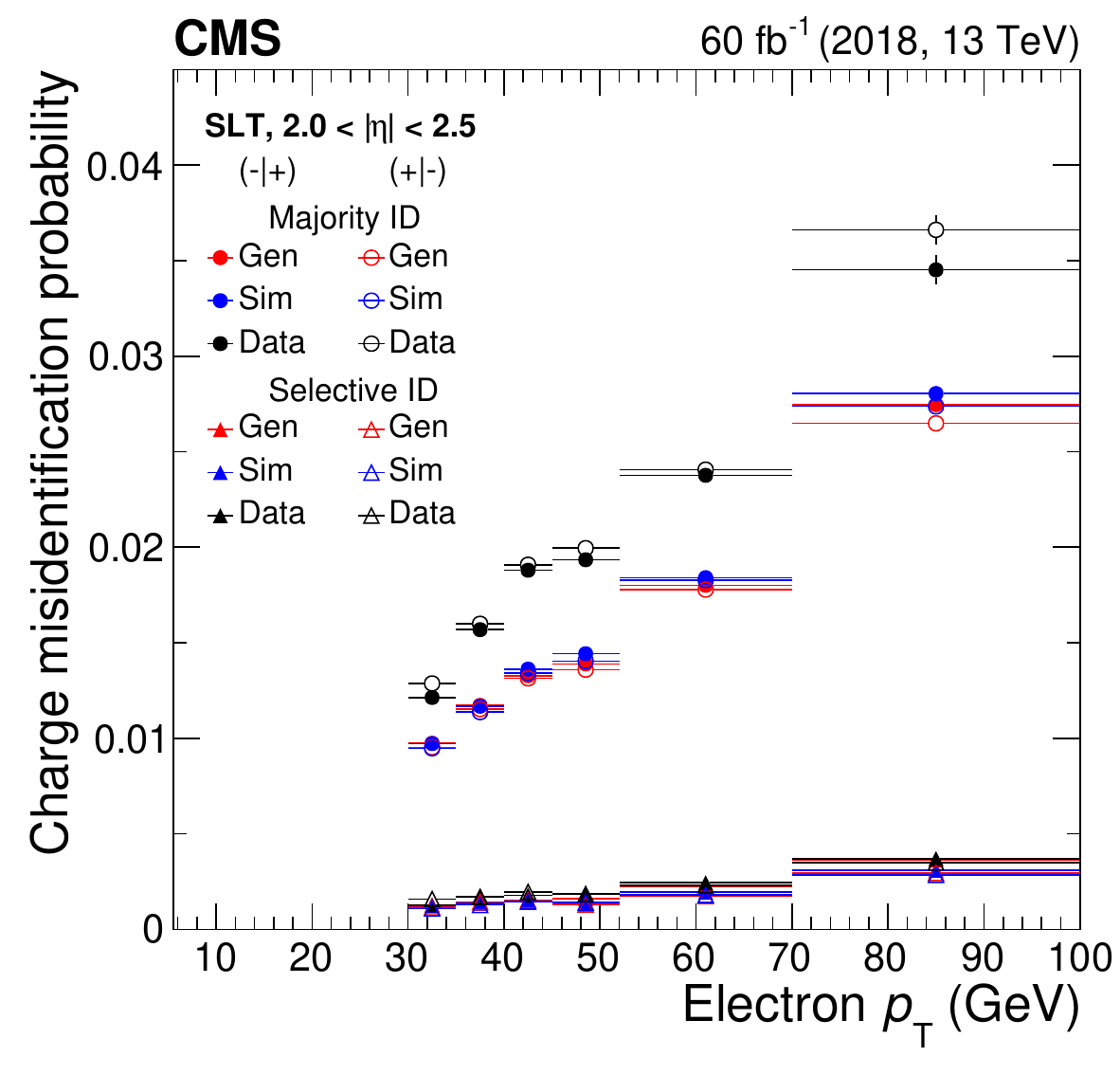}
    \caption{Misidentification rates in the 2018 samples, 
    for electrons in the $2.0 < \feta < 2.5$ bin that pass the single-lepton trigger~(SLT):
    (1)~majority (circles) and selective (triangles) charge identification;
    (2)~misidentification of electrons as positrons~$(+|-)$ (open markers) 
    or positrons as electrons~$(-|+)$ (solid markers);
    (3)~true (red), simulation (blue), and data (black).
    The true charge misidentification rate is evaluated by counting electrons 
    with wrong reconstructed charge using generation-level information;
    the simulated misidentification rate is evaluated with the method used in data.}
    \label{fig_misid}
\end{figure}

The small electron charge misidentification rates
for the ``selective" and ``majority" charge-identifications~\cite{CMS:2020uim} 
are evaluated as a function of the electron \pt and \feta, with a maximum-likelihood fit, 
in the measured and simulated samples,
using same-sign and opposite-sign dielectrons (at the \PZ peak).
The electrons used for this purpose are also categorized depending on 
whether they pass the single-lepton trigger selection. 
Figure~\ref{fig_misid} shows the misidentification rates for several configurations, 
using the 2018 samples for illustration.
The misidentification rates in simulation (``Sim")  agree well with the values obtained
by counting the electrons with misidentified charge according to the generator information (``Gen").
As expected, the selective charge identification option leads to smaller misidentification probabilities
than the majority option.
The probabilities of misidentifying electrons as positrons and vice-versa are equal within uncertainties, 
for all data-taking periods.
The charge misidentification rates in data and simulation are used to correct the simulation.
The full correction, which is used as a conservative estimate of its uncertainty, has a negligible impact on this measurement. 
For dimuons, the charge misidentification is negligible~\cite{CMS:2021yvr}.

We also correct the residual mismodeling of the dilepton \pt distributions
with weights derived iteratively and applied to the pre-FSR simulated dilepton \pt, 
in bins of rapidity. 
The uncertainty associated with this correction is small and 
the full correction is assigned as a systematic uncertainty.

The lepton momentum scale and resolution are calibrated using 
the dilepton mass distributions in Drell--Yan events.
The detailed procedures and parameterizations of the scale and resolution corrections depend on the lepton,
but the general approach is similar for the four lepton categories~\cite{Bodek:2012id}. 
The corrections are applied to both the measured and simulated leptons momenta, 
so that the average dilepton mass values in a narrow window around the \PZ resonance, 
in various bins of kinematic variables, match the corresponding reference values,
derived from the smeared momenta of post-FSR muons or dressed generated electrons 
(produced by combining the momenta of post-FSR electrons and nearby photons within a $\Delta R=0.1$ cone).
Also the resolution of the simulated events matches that of the data;
they are first parameterized using the true lepton-momentum response distributions 
and then corrected by fitting the dilepton mass distributions.

\section{Backgrounds}
\label{sec_backgrounds}

\begin{figure}[htb]
\centering
\includegraphics[width=\cmsFigWidth]{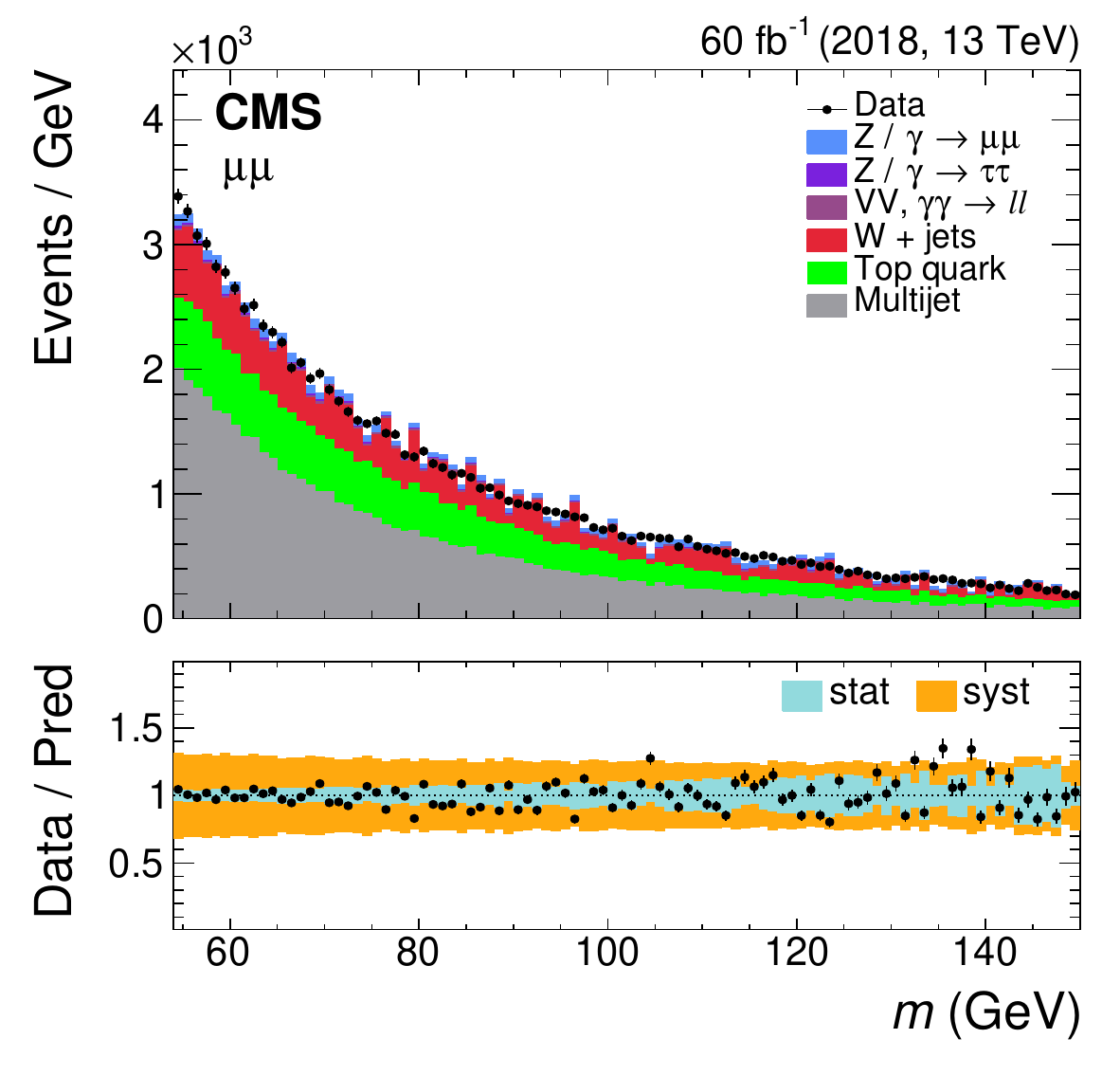}
\caption{Same-sign dimuon mass distribution for the 2018 sample.
The EW and top quark backgrounds are normalized 
to the integrated luminosity using NNLO cross sections.
The multijet background is evaluated by applying weights
to the corresponding multijet-enriched samples.
The error bars in the lower panel include statistical and background systematic uncertainties 
(described in Section~\ref{sec_syst}).}
\label{fig_ssll}
\end{figure}

The small multijet background in the signal region 
is evaluated from multijet enriched control regions (CRs) in data,
using transfer factors (TFs) 
that depend on the muon or electron \feta, \pt, and category.
Four exclusive categories are considered:
leptons that pass single-lepton triggers,
leading and trailing leptons that pass dilepton triggers, 
and nontriggering leptons.
To select the CRs, the following inverted lepton identification criteria are used:
muons must fail the isolation criteria and electrons of all categories (central, forward ECAL, and HCAL) must fail the loose ID criteria 
(which has an efficiency of about 90\%). Central electrons must additionally fail at least two criteria from the medium ID.
As shown in Fig.~\ref{fig_ssll}, the data and predictions agree in the same-sign multijet CR, 
which validates the multijet background evaluation method.

\begin{figure}[ht]
\centering
\includegraphics[width=\cmsFigWidth]{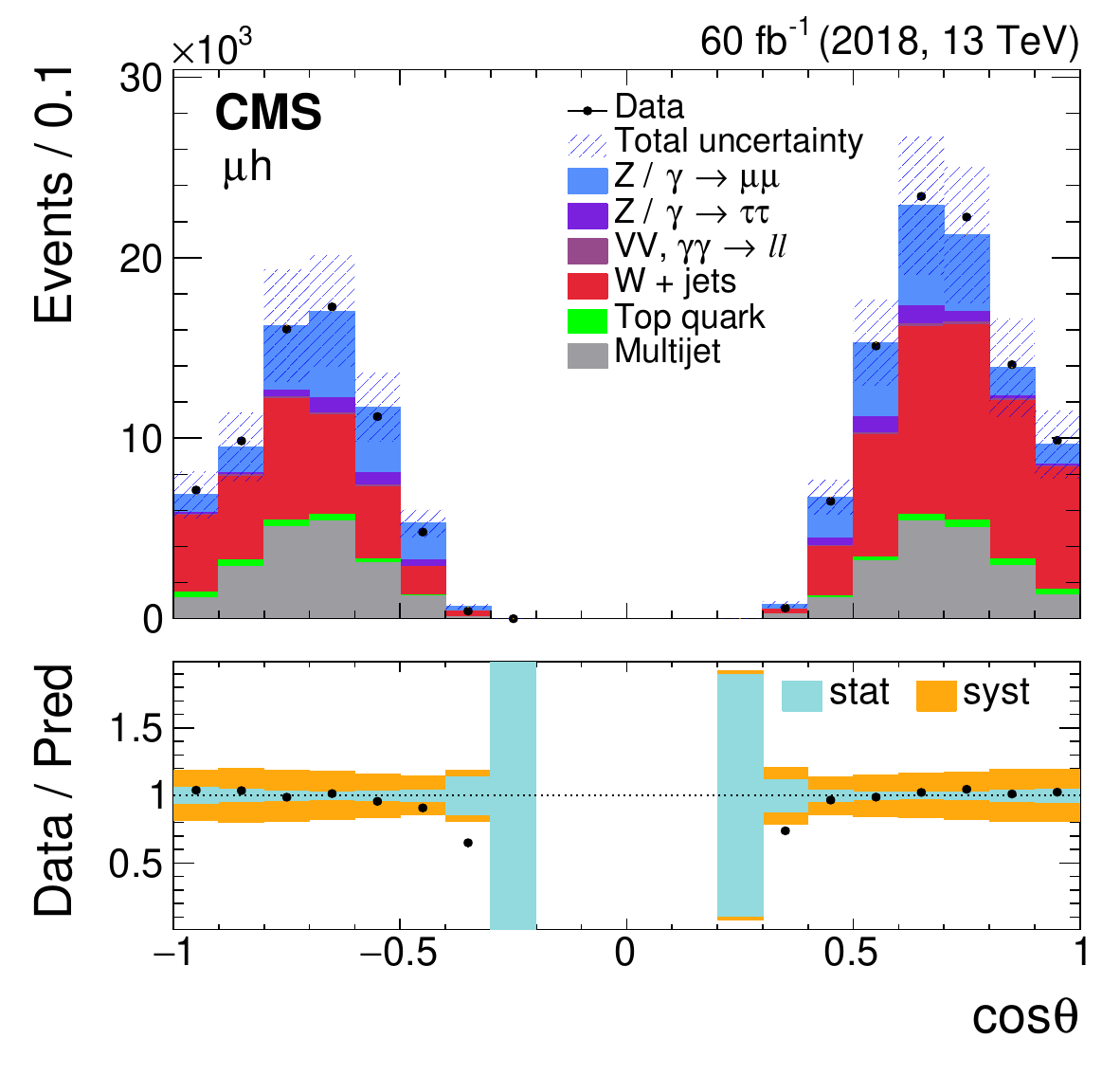}
\caption{Lepton \cs distribution in \uh events in 2018.
The multijet and \wjet backgrounds are scaled to the data as described in the text. 
The error bars include statistical and background systematic uncertainties (described in Section~\ref{sec_syst}).}
\label{fig_uh}
\end{figure}

The normalization of the EW and top quark backgrounds
to the integrated luminosity uses NNLO inclusive cross sections and is validated
(and the corresponding systematic uncertainties are evaluated)
using \ue samples selected using single-muon or single-electron triggers. 
The central-forward $\PGm\Pg$ and $\PGm\Ph$ control regions 
are used to validate the multijet background evaluation and
derive residual \feta- and \pt-dependent scale factors 
to be applied to the simulated \wjet events with incorrectly identified forward electrons.
The scale factors are derived in two steps: 
first, the inclusive transverse mass (\mt) distribution is fitted with floating multijet and \wjet contributions;
then, \feta- and \pt-dependent scale factors are derived from the \wjet enriched CR, 
selected with an additional $\mt > 40$\GeV requirement.
In the \wjet-enriched region, the multijet background estimate is scaled using the results from the first step. 
Figure~\ref{fig_uh} compares the measured and simulated \cs distributions, after applying the scale factors. 
In the CF channels with a forward HCAL electron, events with small $\abs{\cs}$ values do not pass the acceptance cuts.

\section{Systematic uncertainties}
\label{sec_syst}

Besides the statistical uncertainties in the lepton efficiencies, 
we consider 31 independent sources of systematic uncertainty affecting the efficiency for central leptons, 
including
signal shape model,
background model,
tag selection,
pileup model,
trigger prefiring probabilities,
dilepton \pt model,
beamspot position, 
as well as mass range and bin width of the dilepton mass histograms used in the tag-and-probe method.
We also consider the
generator-matching requirement used to make a signal template,
fitting instead of counting in the determination of the trigger efficiencies,
alternative background subtraction (only for electrons),
alternative factorization of reconstruction times identification efficiencies (only for muons), 
and a residual $\eta$-dependent scaling factor
that makes the simulated lepton $\eta$ distribution agree with the data.

For the \Pg and \Ph electrons, whose efficiency measurements~(especially for \Ph) are more challenging because the signal 
and background dilepton mass shapes are more similar in some measurement bins, 
we consider fewer but more conservative systematic variations.
The 9 systematic sources include 
the signal shape model,
3 alternative variations in the event selections to change the signal and background contributions~(2 for the central tag electron, 1 for the \mt),
fit range,
dilepton \pt model, 
trigger prefiring probabilities,
beamspot position,
and dilepton mass window for the integral calculation.

For the multijet background, a conservative $\pm50\%$ uncertainty covers the residual 
differences seen in the \ug and \uh control regions.
The \wjet background shapes are obtained from simulation, 
but in the forward channels scale factors are applied 
as functions of the forward,
often misidentified,
electron \pt and \feta.
A 25\% uncertainty is estimated by varying the multijet contribution by 50\%
in the \wjet CR where the scale factors are evaluated.
For the central-central channels, both the multijet and \wjet contributions are very small.
An uncertainty of 10\% is assigned to the top quark and $\ztt$ background estimates,
consistent with the differences observed in the \ue CR.
Other backgrounds have negligible uncertainties.

The statistical uncertainties in the lepton momentum corrections 
are evaluated by generating bootstrap replicas~\cite{efron1979} for data and simulated samples.
We consider the following systematic sources:
background estimate, 
dilepton \pt model, 
mass ranges for scale and resolution corrections, 
a material correction bias for muons, 
a nonlinearity correction to the electron energy,
and different parameterizations of resolution corrections for central and forward electrons. 

In addition to the statistical uncertainties in the data sample 
used to determine the weight factors applied to the simulated samples to correct for the trigger prefiring,
we consider systematic uncertainties of $\pm20\%$ in each measurement \feta bin in two ways:
independent of \pt and linear versus $1/\pt$ for $\pt > 20$\GeV. 
We also vary the trigger prefiring probability of each object linearly with the event number in each run
(\ie, approximately versus time)
so that the average measured per-object trigger prefiring probability is unchanged.
The ECAL trigger prefiring probabilities, measured in bins of $\eta$ and \pt, are parameterized
as functions of \pt in each $\eta$ bin. 
The differences between the best fit values and the per-bin measurements 
are an additional systematic uncertainty.

To estimate the uncertainty associated with mismodeling of the beamspot position, 
we reweight the MC samples 
using the measured distribution of the primary vertex $z$ position.
Pileup uncertainties are estimated varying by $\pm 5\%$ the nominal total inelastic cross section~\cite{Sirunyan:2303548}
used to calculate the pileup distribution. 
The electron charge misidentification uncertainty is evaluated
by removing from the simulation the correction to the misidentification rate.

The renormalization and factorization scales, $\mu_\mathrm{R}$ and $\mu_\mathrm{F}$, 
are each varied independently by a factor of 2, 
up and down, such that their ratio remains within $0.5 < \mu_\mathrm{R} / \mu_\mathrm{F} < 2.0$. 
The maximum deviation among these six variants relative to the nominal choice 
is assigned as a systematic uncertainty associated with the missing higher-order QCD corrections
and is included (in quadrature) in the theory uncertainty. 
The uncertainty in the modeling of the dilepton \pt distribution is estimated
as the difference between correcting or not the simulated \pt distribution to match the data.

The following uncertainties, due to missing higher-order EW corrections, are considered:
the treatment of finite-width effects 
by comparing the default complex mass (CM) scheme with the pole scheme (PS) for the propagator;
and the difference between the default EW input scheme ($G_\mu$, \sineff, $m_\PZ$) 
and an alternative scheme using the $\alpha_\mathrm{em}$, \sineff, and $m_\PZ$ parameters.
We also evaluate the effect of uncertainties in the input parameters of the default scheme,
by varying $G_\mu$ and the masses of the \PZ boson and of the top quark
within their uncertainties~\cite{Workman:2022ynf}.
The uncertainty affecting the FSR modeling is estimated in two ways:
as the difference between the \PHOTOS values including or not
matrix element corrections to the first emission,
and as the difference between the \PHOTOS and \PYTHIA evaluations.
Uncertainties due to the missing effects of interference between the initial and final state 
radiation are considered negligible.

The PDFs dominate the uncertainty of \sineff measurements in hadron collisions because
they affect the size of the dilution of the observed \afb 
and the fraction of dileptons produced in in \PQu\PAQu and \PQd\PAQd
interactions. Several PDF sets, described in the next sections, are considered to evaluate the corresponding uncertainty. 
Using the fact that \sineff and the PDFs affect differently the $m_{\Pell\Pell}$-dependence of the observed \afb, 
the PDF uncertainties are profiled~\cite{Bodek:2015ljm} while extracting \sineff. 
For profiling, we use the nominal tolerance criterion of $\Delta\chi^2=1$.

\begin{figure*}[ht]
\centering
\includegraphics[width=0.329\textwidth]{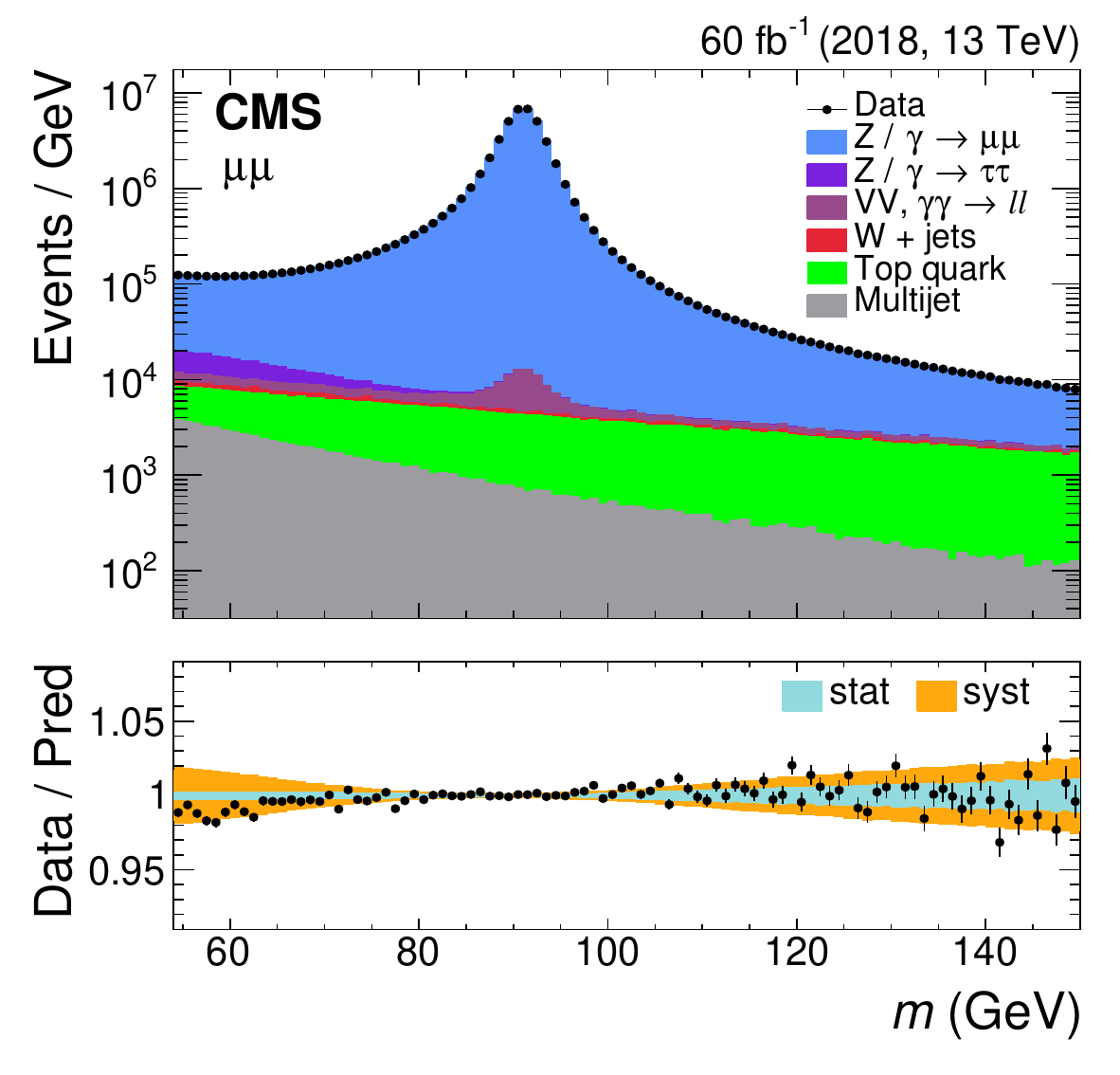}
\includegraphics[width=0.329\textwidth]{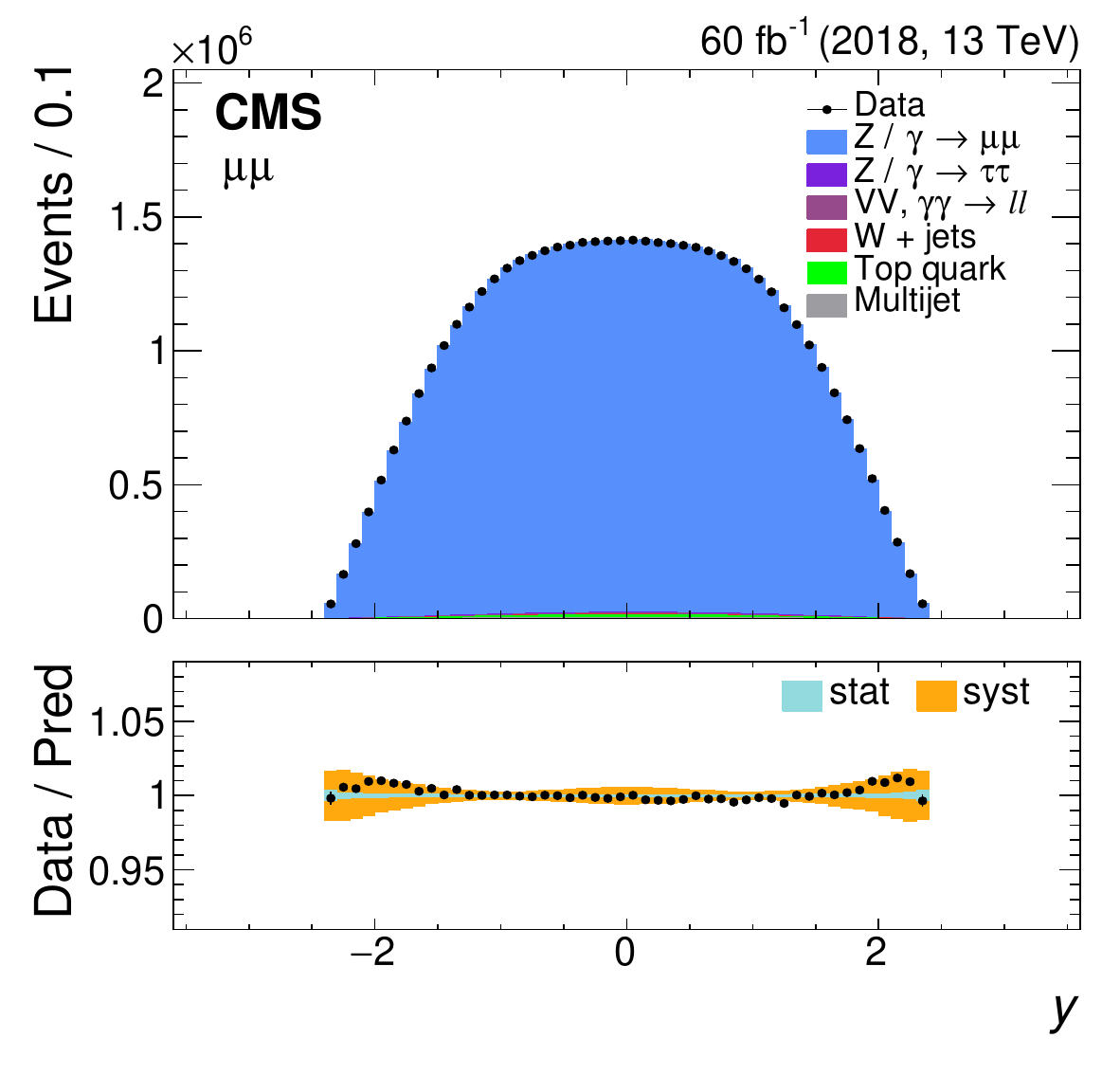}
\includegraphics[width=0.329\textwidth]{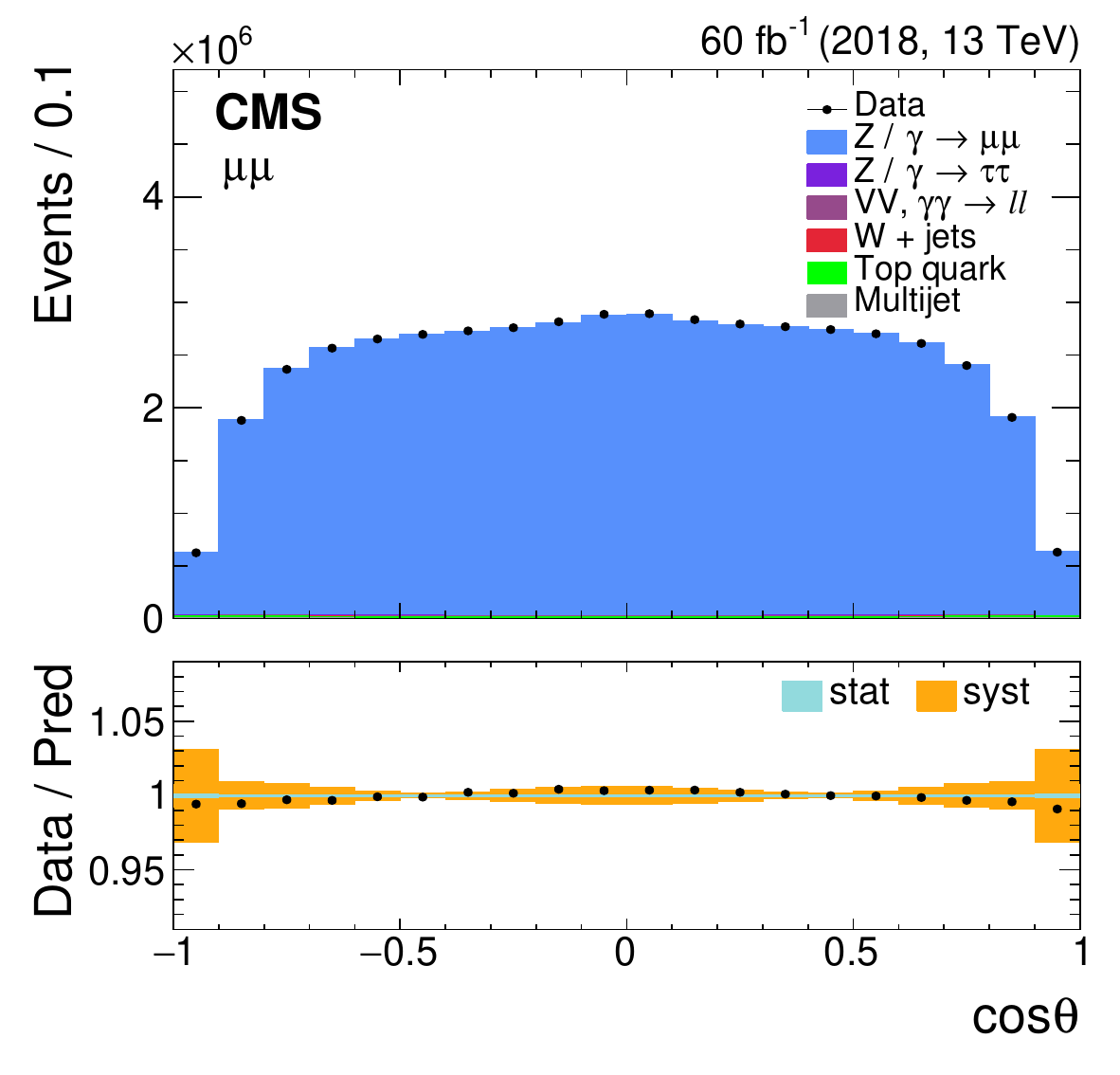}
\includegraphics[width=0.329\textwidth]{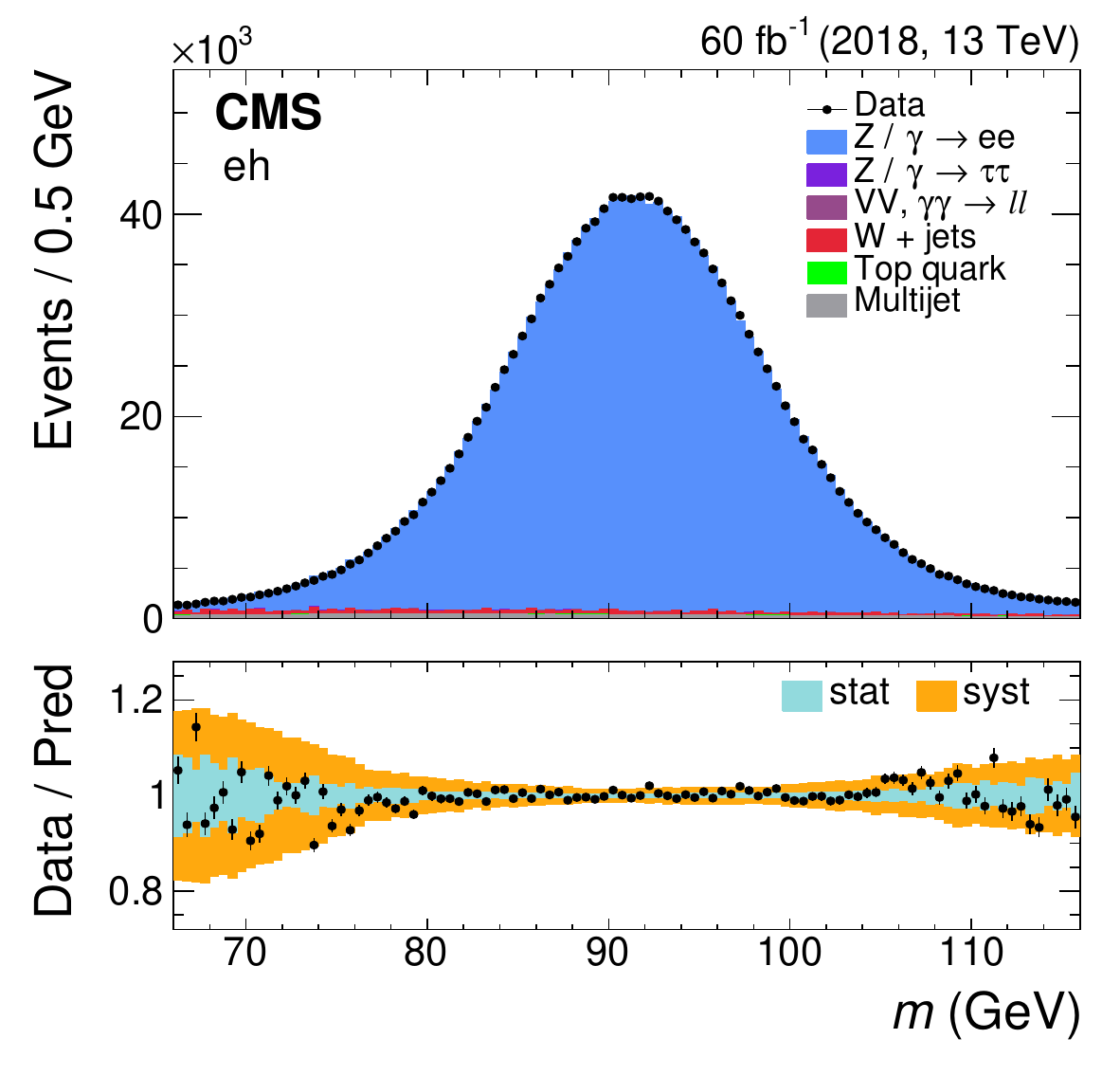}
\includegraphics[width=0.329\textwidth]{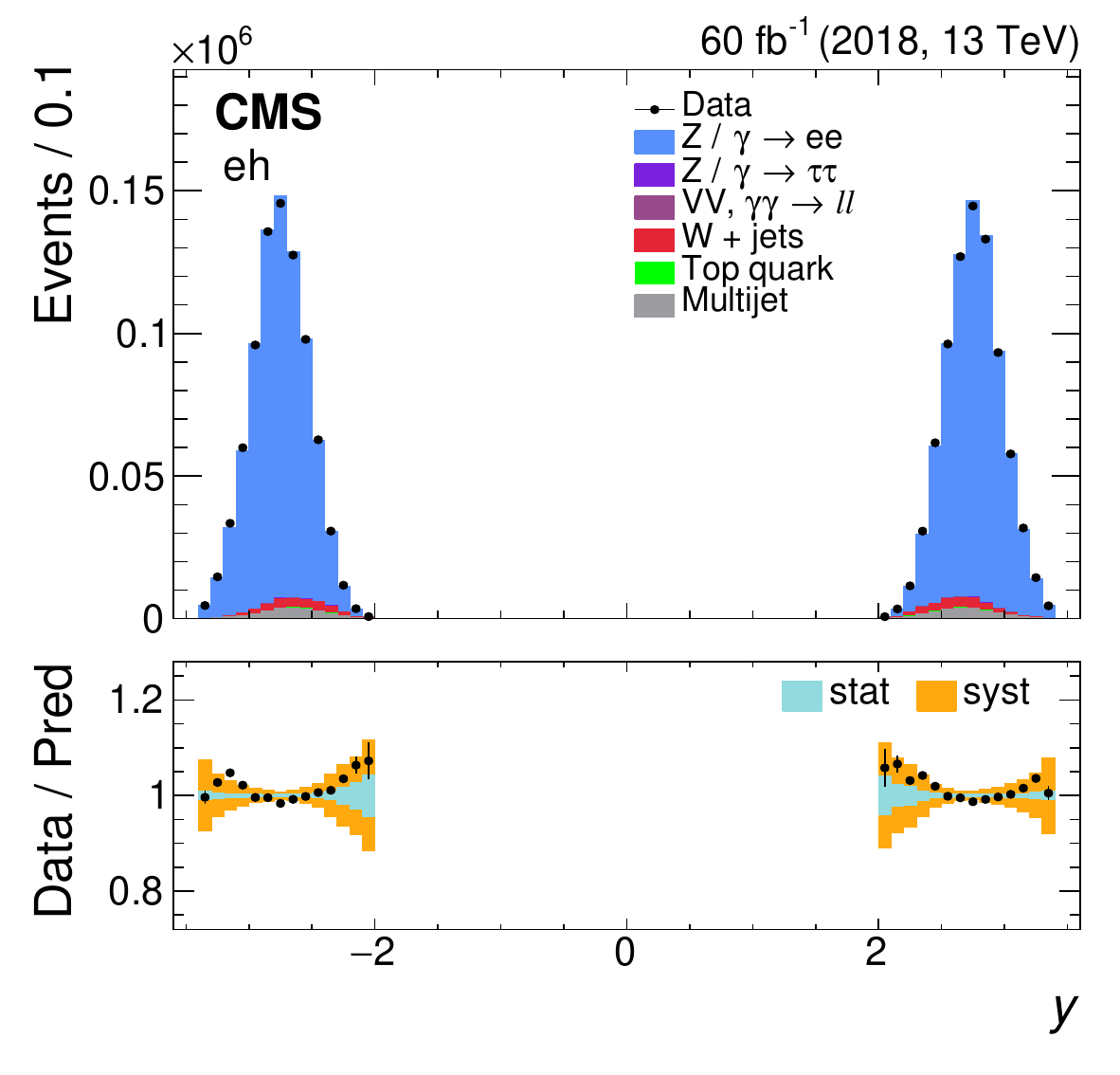}
\includegraphics[width=0.329\textwidth]{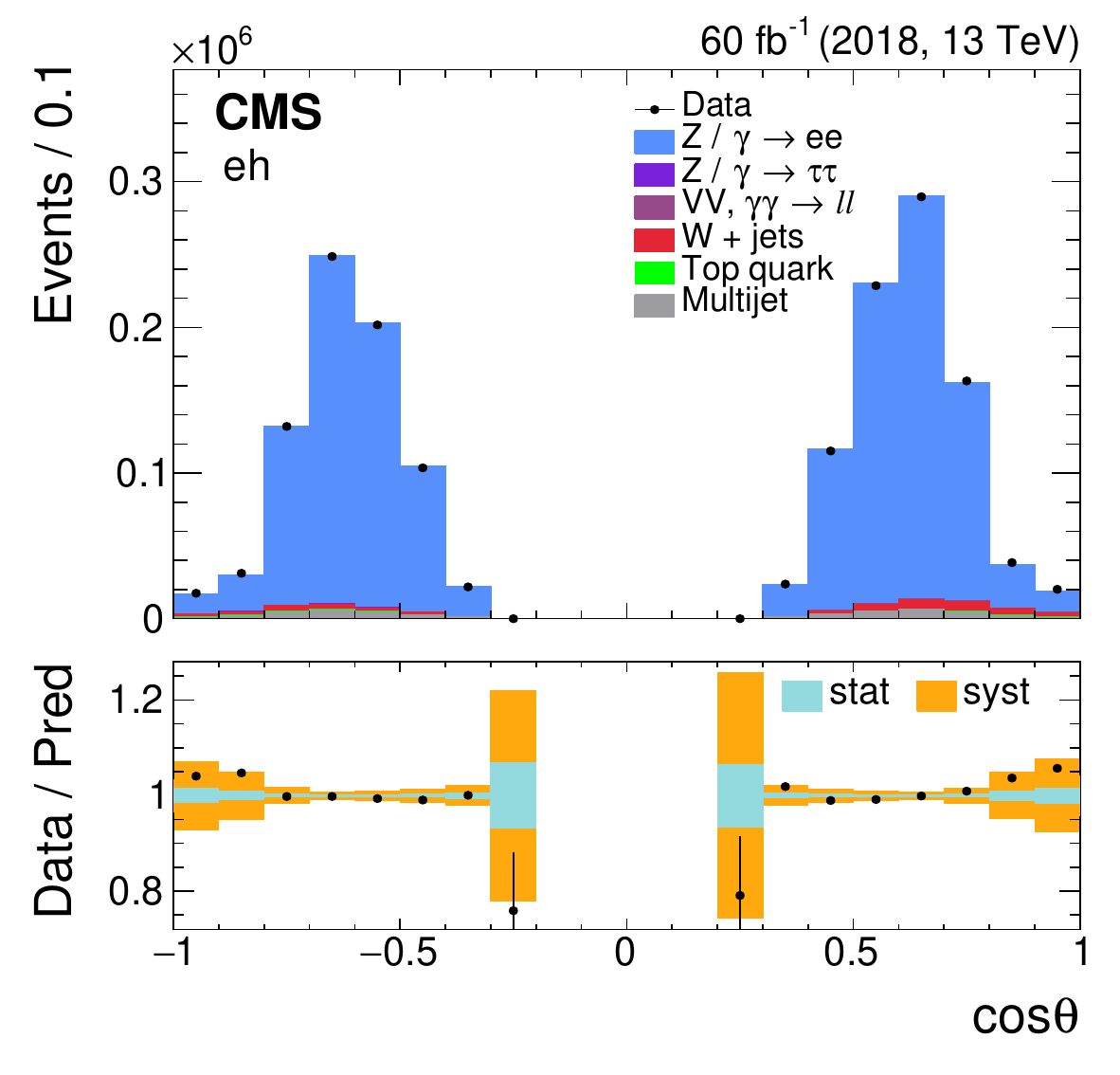}
\caption{Dilepton mass (left), rapidity (middle), and \cs (right) distributions,
for the \uu (upper) and \eh (lower) channels in the 2018 sample, 
after applying all the corrections.
The signal is scaled to match the total number of events in the data.}
\label{fig_llmycs}
\end{figure*}

Figure~\ref{fig_llmycs} compares the measured and predicted distributions of
the dilepton mass, rapidity, and \cs, 
using the \uu and \eh channels in the 2018 data sample.
The simulated and measured distributions agree within their uncertainties.

\section{Measurement bins and interpretation model}
\label{sec:bins}

We measure \sineff in two ways:
by fitting the observed weighted \afbwym distribution and 
by fitting the unfolded angular coefficient \afourym of the pre-FSR dilepton system.
The measurement is restricted to $|y|<3.4$
to avoid regions where the signal acceptance is small.
The mass bins in the \PZ region are narrow 
(but wider than the dimuon mass resolution),
whereas elsewhere they are broad, to compensate for the smaller yields.
Table~\ref{tab_bins} lists the mass and rapidity bins used in the fits.

\begin{table*}[t!] 
\centering
\topcaption{Dilepton rapidity and mass binning used in the fits; 
$n_{\abs{y}}$, $n_m$, and $n_M$ are the numbers of bins in each category.}
\label{tab_bins}
\begin{tabular}{ c  c  c  c  c  c  c  c  c  c  c  c  c   c }
Channel& \multicolumn{12}{c}{Absolute rapidity bin edges} &  $n_{\abs{y}}$  \\
\hline
\uu, \ee &  0.0  & 0.4  & 0.8  & 1.2  & 1.6  & 2.0  & 2.4 &   &  &  &   &  &  6 \\
\myeg &  &  &  & 1.2  & 1.6  & 2.0  & 2.4  & 2.7  &  &  &   &  &  4 \\
\eh &  &  &  &  &  &  2.0  & 2.4  & 2.7   & 3.0  & 3.4   &   &  &  4 \\[1mm]
\multicolumn{14}{c}{} \\ [-1.8ex]
Channel &\multicolumn{12}{c}{Reconstructed $m$ bin edges (\GeVns)  } & $n_m$ \\
\hline
\uu, \ee &  54 & 66 & 76 & 82 & 86 & 89.5 & 92.7 &  96 & 100 & 106 & 116 & 150 & 11 \\
\myeg, \eh &  & 66 & 76 & 82 & 86 & 89.5 & 92.7 &  96 & 100 & 106 & 116   &   & 9 \\[1mm]
\multicolumn{14}{c}{} \\ [-1.8ex]
$\abs{Y}$ range & \multicolumn{12}{c}{Pre-FSR $M$ bin edges (\GeVns)} & $n_M$ \\ \hline
0.0--1.2  &  54 & 66 & 76 & 82 & 86 & 89.5 & 92.7 &  96 & 100 & 106 & 116 & 150 & 11 \\
1.2--2.4  &  54 & 66 & 76 &      & 86 &         &         &  96 &        & 106 & 116 & 150 & 7 \\
2.4--3.4  &      &  66 &      &      & 86 &         &         &  96 &        &        & 116 &        & 3 \\
\end{tabular}
\end{table*}

To construct the template model used in the fits, in both methods, 
we use two different \POWHEG event generator programs: ``\minnlo" and ``\textsc{z}\_ew".
The full-simulation events 
(about 1.4 billion, adding all data-taking periods and channels) 
are generated with the \POWHEG \minnlo generator.
They are complemented by almost 10 billion \minnlo generator-level events, of identical settings,
to smooth the templates used in the reconstructed \afbw fits and in the interpretation fits, 
as well as by 0.3 billion events with alternative PDF weights.
We also generated several samples with \powhegzew,
which provides NLO+HO weak corrections and supports 
a \sineff EW input scheme at NLO in EW~\cite{Chiesa:2019nqb},
to construct templates for \sineff variations, 
as well as for systematic variations in the electroweak input schemes, 
input parameters, and alternative FSR models.

The default interpretation model is constructed by 
reweighting the \POWHEG \minnlo events with: 
(1)~LO-to-NLO$+$HO weak correction factors, evaluated with ``\textsc{z}\_ew", that correct 
the cross sections and all angular coefficients in pre-FSR bins of $\abs{Y}$, $M$, and $P_\mathrm{T}$;
(2)~$\abs{Y}$-dependent $P_\mathrm{T}$ weights that correct the corresponding reconstructed \pt distributions to data; and
(3)~smoothing weights to match the $\afour(\abs{Y}, M, P_\mathrm{T})$ distribution 
of the large \POWHEG \minnlo sample at the generator level in full phase space.
The smoothing weights greatly reduce the statistical fluctuations in pre-FSR distributions and also, to a lesser extent,
in the reconstructed distributions in the smaller full simulation sample.

\begin{figure}[t!]
\centering
\includegraphics[width=\cmsFigWidth]{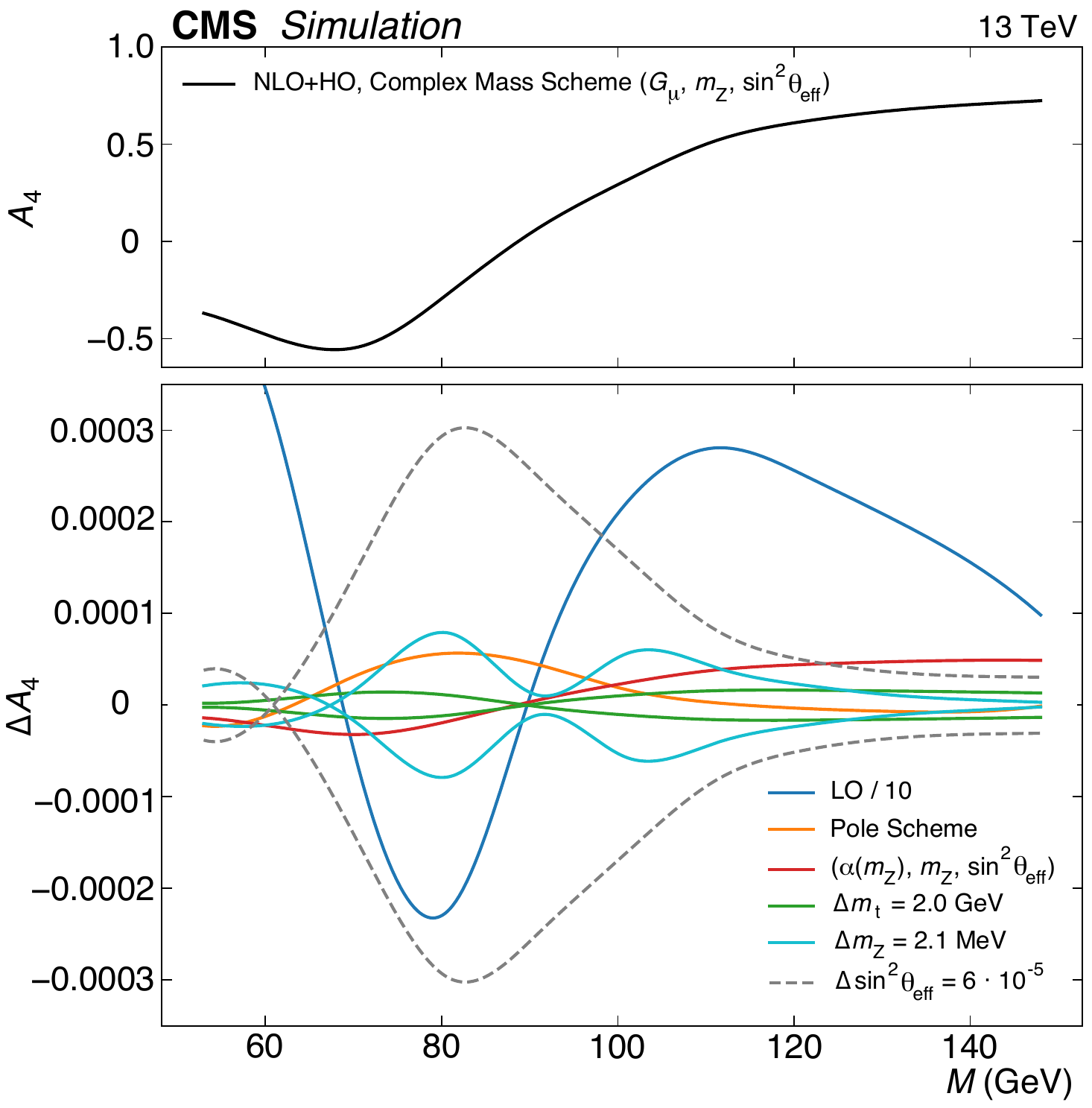}
\includegraphics[width=\cmsFigWidth]{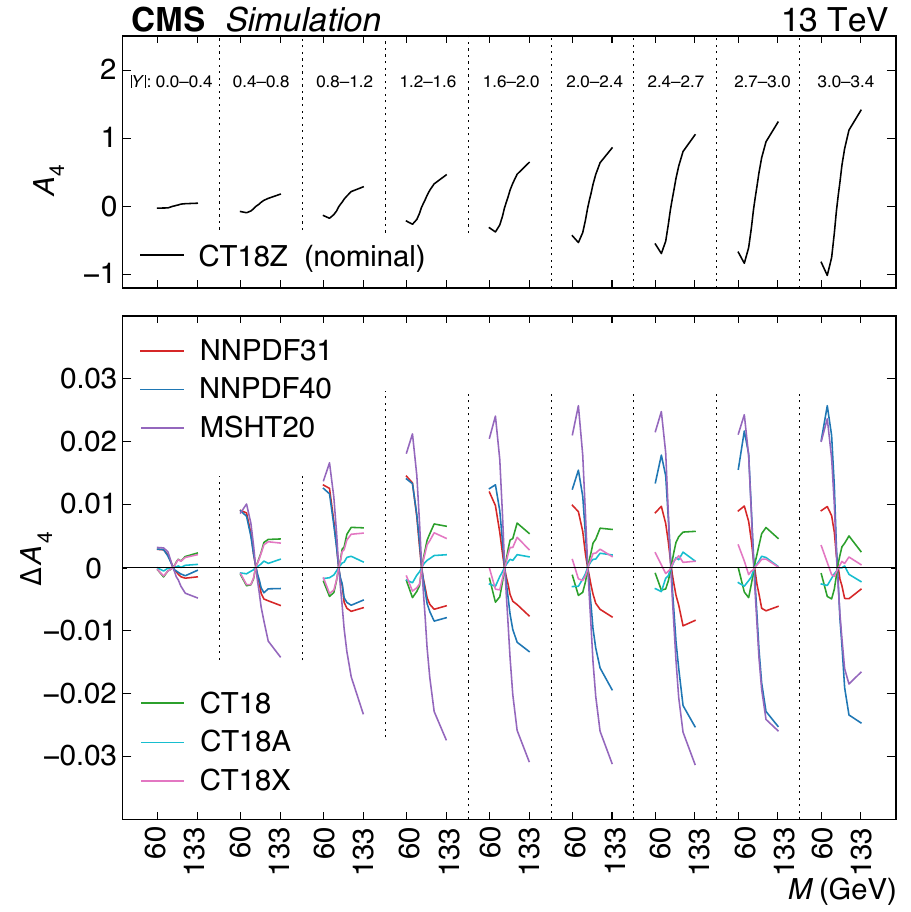}
\caption{The $\afour$ coefficient in the nominal configuration (upper panels) 
and its variations (lower panels) when changing the inputs mentioned in the legends:
different \powhegzew options~(\cmsLeft) and different PDF sets~(\cmsRight).
No lepton kinematic selection criteria are applied. 
The size of the \sineff variation is chosen for a clear visual demonstration.}
\label{fig_model}
\end{figure}

The FSR variations are applied using mass and rapidity distributions of the alternative FSR samples. 
The variations in the \sineff, PDFs, $\mu_R$, and $\mu_F$ input parameters 
are based on the corresponding samples, 
and per-event weights are built using the dilepton $\abs{Y}$-$M$-$P_\mathrm{T}$ 
triple-differential cross sections and angular coefficients. 
Figure~\ref{fig_model} shows \afour for different electroweak configurations
and PDF sets.

All the fits described in the following sections were blinded by shifting the fitted \sineff by a random offset 
until all the systematic uncertainties were finalized and cross-checks completed.

\section{Extraction of \texorpdfstring{\sineff}{theta(eff) } from \texorpdfstring{\afbw}{AFB}}
We extract \sineff by fitting the \afbwym angular-weighted asymmetry 
to templates varying \sineff and PDF set, 
similarly to the analysis performed with the $\sqrt{s} = 8$\TeV data~\cite{cms:wma}.
This method uses reconstructed variables; we use it to provide our baseline result because 
the positive and negative \cs have approximately identical detection acceptance and efficiencies, 
so that their systematic uncertainties cancel to a large extent~\cite{Bodek:2010qg}.
The dependence of \afbw on the dilepton mass constrains the PDFs 
while extracting \sineff through a so-called profiling procedure~\cite{Bodek:2015ljm}.
As described in Ref.~\cite{Bodek:2010qg}, the definition of \afbwym depends on $A_0$ and, in the following, 
we use the simple tree-level expression for $q \overline{q}$ events, $A_0 = \pt^2 /(\pt^2 + m^2)$. 
Almost identical results are obtained if we use, instead,
a fixed average $A_0$ value in each $y$-$m$ bin
or an empirical $A_0$ parameterization based on simulated samples.

The \sineff value is found by minimizing
\begin{equation*}
\chi^2(s,\vec{\theta}) = |\vec{\theta}|^2 + \sum_{i} \left(D_i-T_i(s,\vec{\theta})\right)^T V_i^{-1}\left(D_i-T_i(s,\vec{\theta})\right),
\end{equation*}
where $s \equiv \sineff$,
$i$ represents the data-taking periods and dilepton channels,
$D$ is the measured angular-weighted \afbwym with $n_m \times n_{\abs{y}}$ bins,
and $V$ is the covariance matrix, 
which includes the statistical uncertainties in data and templates,
as well as in the lepton calibration, efficiencies, and trigger prefiring probabilities.
The predicted \afbwym, denoted by $T(s,\vec{\theta})$,
includes background contamination and is obtained by combining all the individual 
differences corresponding to either $s$ or $\theta_i$ with respect to their nominal values.
Lastly, $\vec{\theta}$ is the vector of nuisance parameters, 
including the experimental systematic uncertainties and the PDFs.
The CT18Z PDF set~\cite{Hou:2019efy} was selected 
(before unblinding the analysis) to report the nominal result 
because it provides an accurate description of the CMS data and offers, 
within its uncertainties, the best coverage of the central values obtained with the other PDF sets
(as will be shown later).
Furthermore, this PDF set imposes identical strange quark and antiquark densities in the proton,
a constraint supported by 
measurements of \PW boson production with associated charm quarks~\cite{CMS:2018dxg,CMS:2013pzl}
that is not assumed in the MSHT and NNPDF sets.

\begin{table}[!t] 
\centering
\topcaption{Free parameters in the \afbwym fit, 
indicating the number of independent variations 
(\eg, number of rapidity bins where the uncertainties are considered uncorrelated).
Some of the total
values reflect the four data-taking periods and/or the four final-state channels.}
\label{tab_parameters}
\cmsTable{
\begin{tabular}{ l l c c } 
Category & Parameters              &  Variations  & Total \\ \hline
             &  \sineff                                &    1      &   1  \\ 
PDF     &  CT18Z                               &   29     &  29  \\
theo     &  \pt model ($y$)                  &    9     &   9  \\ 
theo     &  FSR model ($\PGm,\Pe$) &  2, 2   &   4  \\ 
eff        &  Muon                                 &   31    & 124  \\
eff        &  Central electron                 &   31    & 124  \\
eff        &  ECAL forward electron      &    9    &  36  \\
eff        &  HCAL forward electron      &    9    &  36  \\ 
calib    &  Muon                                  &    4    &  16  \\
calib    &  Central electron                 &    7     &  28  \\
calib    &  ECAL forward electron      &   11     &  44  \\
calib    &  HCAL forward electron      &    9     &  36  \\ 
bkg      &  Multijet ($y$)                     &  6, 4    &  80  \\
bkg      &  \wjet ($y$)                         &  6, 4    &  80  \\
bkg      &  \ztt                                    &    1      &  16  \\ 
bkg      &  Top quark                         &    1      &  16  \\ 
other    &  Integrated luminosity       &    1      &   4  \\
other    &  Pileup                              &    1      &   4  \\
other    &  Charge misidentification                  &    1      &   4  \\
other    &  Trigger prefiring ($\PGm,\Pe$)    &  4, 4    &  32  \\ 
other    &  Beamspot                        &    1      &  16  \\[\cmsTabSkip]
            &  Total                                 &            & 739  \\
\end{tabular}
}
\end{table}

The nuisance parameters are listed in Table~\ref{tab_parameters},
which reflects various correlations between channels and rapidity bins. 
The experimental systematic uncertainties are treated as uncorrelated 
between the different run periods, while the theoretical uncertainties are treated as correlated. 
Such a correlation scheme, together with the conservative treatment of experimental uncertainties per data-taking period, 
provides flexibility to account for possible differences in the systematic effects originating 
from different data-taking and reconstruction conditions. 
This also allows us to have a more similar correlation scheme with the unfolding procedure, described in the next section, 
for which decorrelations are more important to improve the fit quality
(a fully identical correlation scheme for all parameters is not needed here given that 
\afbw is much less sensitive to many systematic sources and too many free parameters add unnecessary complication 
since the \afbw fits use numerical methods of {\sc Minuit}~\cite{James:1975dr} and are relatively slow). 

\begin{figure}[t!]
\centering
\includegraphics[width=\cmsFigWidth]{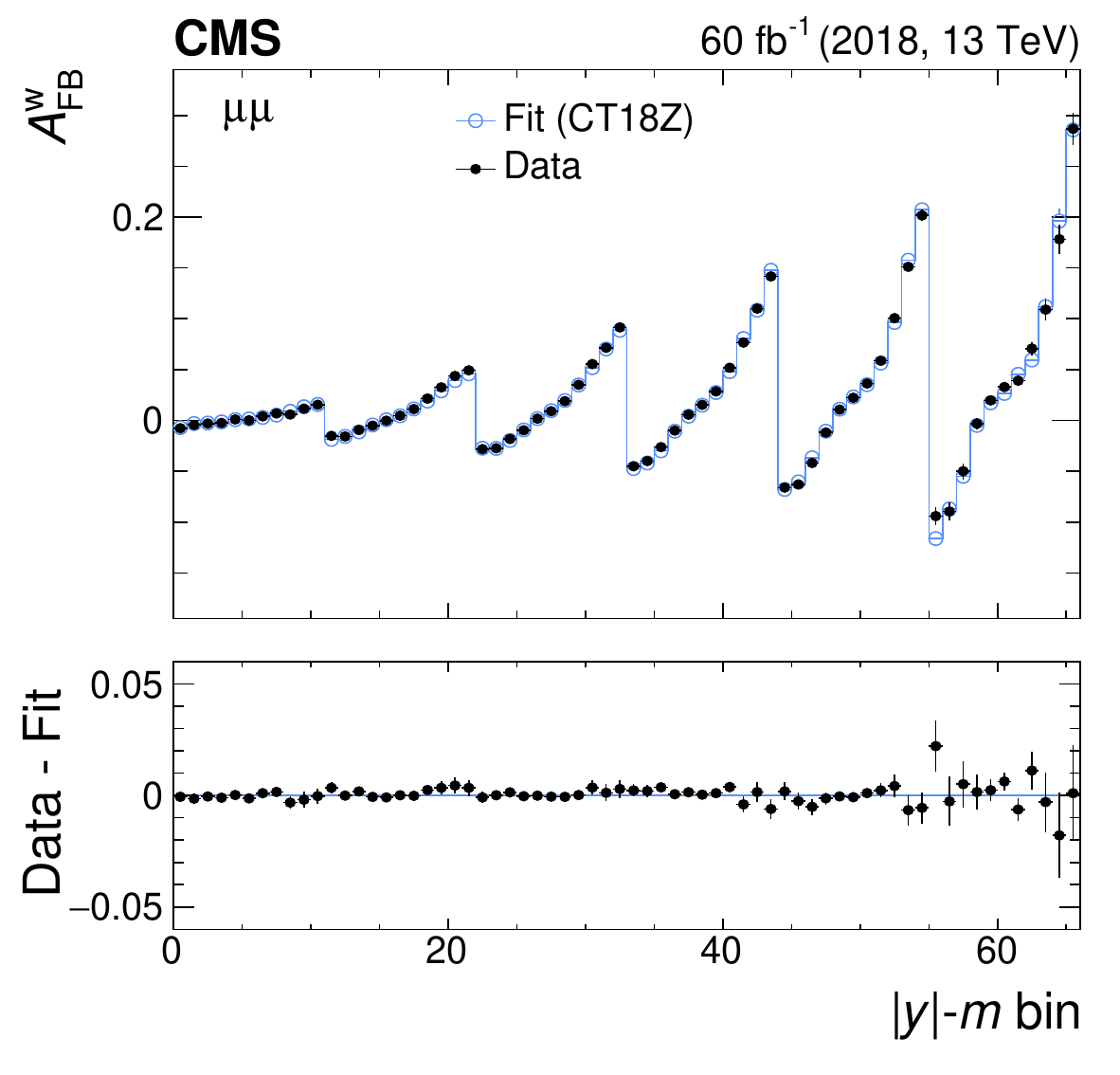}
\includegraphics[width=\cmsFigWidth]{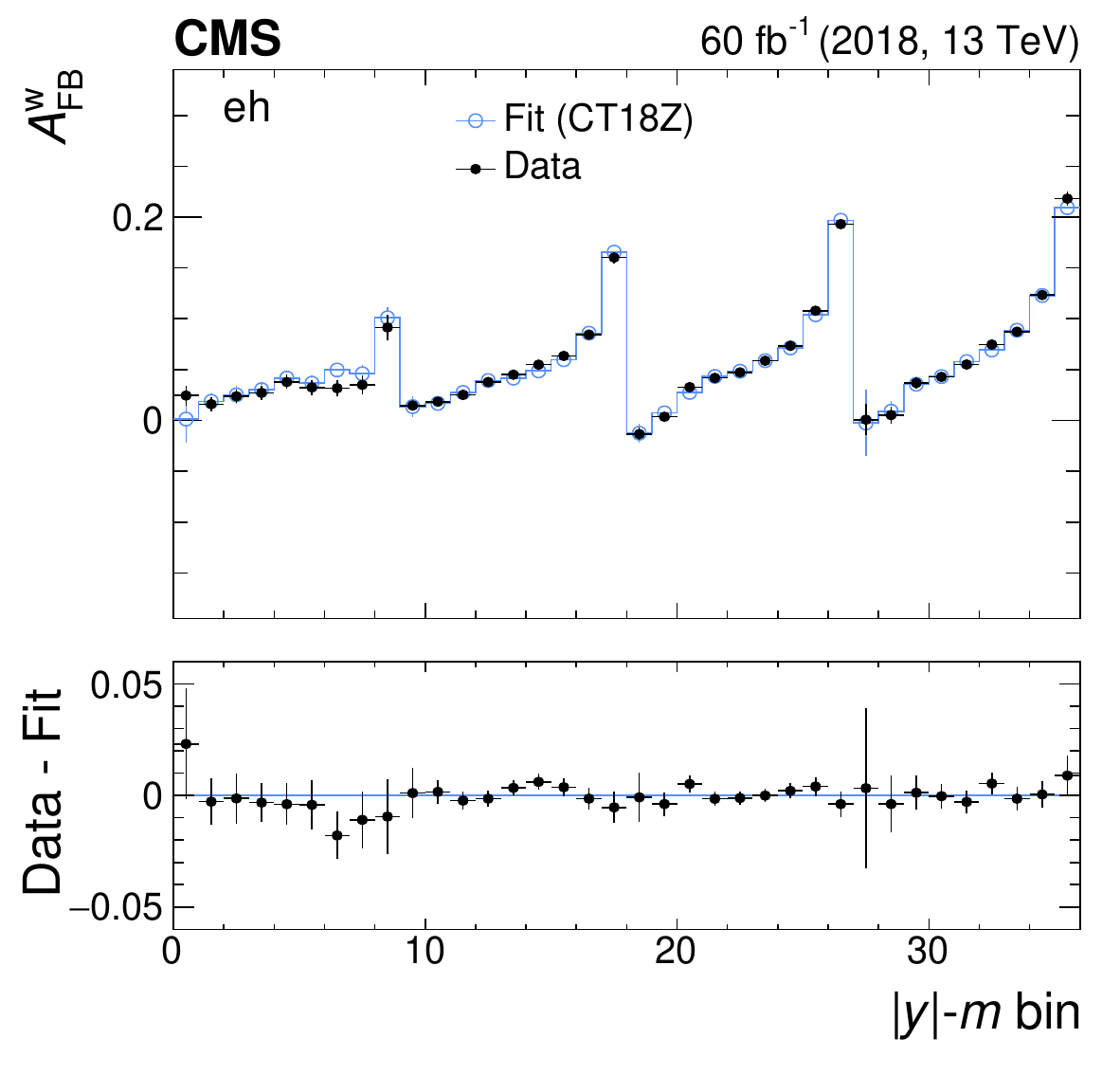}
\caption{Measured and best fit \uu~(\cmsLeft) and \eh~(\cmsRight) 
\afbwym distributions for the 2018 data.
The error bars represent the statistical uncertainties in the measured and simulated samples. 
The rapidity bins are given in Table~\ref{tab_bins}.}
\label{fig_afb}
\end{figure}

The nominal ``combined fit configuration", 
which includes all data-taking periods, all channels, 
the CT18Z PDF set, 
and all other nuisance parameters for systematic uncertainties,
has 739 free parameters.
The fit does not include, as nuisance parameters, 
the renormalization and factorization scale variations, 
nor the EW systematic uncertainty variations. 
Instead, individual fits are performed for each of those options
and the deviations from the nominal result are considered as uncertainties.
Figure~\ref{fig_afb} shows the fitted \afbw distributions 
for the \uu and \eh channels, using the 2018 data sample.
The agreement between the data and best-fit distributions is excellent 
in all channels and years. 
None of the nuisance parameters has a significant pull zero. 
The largest 14 impacts come from the nuisance parameters of PDF uncertainties, 
which dominate the uncertainties in this measurement. 

To estimate the statistical contributions from various experimental systematic uncertainties in \sineff, 
we compute by how much the best fit uncertainty increases when adding 
the corresponding contributions
to the covariance matrix of the statistical uncertainties in the data. 
Individual or grouped systematic uncertainties 
are calculated by fixing the corresponding nuisance parameter(s) 
to the best fit values obtained in the nominal fit
and evaluating the decrease in the uncertainty:
the quadratic difference from the nominal uncertainty 
is taken as the corresponding uncertainty.
\begin{table*}[!t] 
\centering
\topcaption{Fitted \sineff (in units of $10^{-5}$) for the four channels 
and their sum ($\Pell\Pell$), using the full data sample. 
The fit quality is good, as indicated by the $\chi^2$ probabilities ($p$).
The experimental systematic uncertainties (``exp") 
are the sum of the values in the five rightmost columns,
corresponding to the statistical uncertainties of the MC samples 
and the categories listed in Table~\ref{tab_parameters}.}
\label{table_r2_ll}
\begin{tabular}{l c c c c  c c c c  c c c c c}
 &  $\chi^2_\text{min}$ & bins & $p$(\%) & \sineff  & stat & exp & th & PDF & MC & bg & eff & calib & other \\
\hline
\uu          & 241 & 264 & 83 & $23\,142 \pm 38$ & 17 & 17 & 6 & 30 & 13 &  3 &  2 &  5 & 4 \\
\ee          & 257 & 264 & 60 & $23\,171 \pm 41$ & 22 & 18 & 5 & 30 & 14 &  4 &  5 &  3 & 7 \\
\myeg        & 119 & 144 & 93 & $23\,253 \pm 61$ & 30 & 40 & 3 & 44 & 23 & 11 & 12 & 20 & 8 \\
\eh          & 105 & 144 & 99 & $23\,114 \pm 48$ & 18 & 33 & 9 & 37 & 14 & 10 & 16 & 18 & 6 \\[\cmsTabSkip]
$\Pell\Pell$ & 731 & 816 & 98 & $23\,152 \pm 31$ & 10 & 15 & 8 & 27 &  8 &  4 &  6 &  6 & 3 \\
\end{tabular}
\end{table*}
Table~\ref{table_r2_ll} presents the results obtained for the four channels 
using the full data sample, specifying the total and partial uncertainties. 
The largest contribution to the total theoretical uncertainty comes from 
the renormalization and factorization scale variations (0.00008).
The four results are in agreement, as expected from lepton flavor universality.
The \uu and \ee channels give a combined uncertainty of 0.00034, which is reduced to 0.00031 
by adding the CF channels.

\section{Measurement of \texorpdfstring{\afour}{A4} and extraction of \texorpdfstring{\sineff}{sin[eff]}}
\label{sec_a4}

We also extract \sineff through the unfolded $\afour$ measurements, at Born level,
in the pre-FSR dilepton $\abs{Y}$-$M$ bins.
The \afourym values  are obtained by minimizing
\begin{equation*}
\chi^2(\vec{p},\vec{\nu}) 
= \abs{\vec{\nu}}^2 + \sum_{i} \left(D_i-T_i(\vec{p},\vec{\nu})\right)^T V_i^{-1}\left(D_i-T_i(\vec{p},\vec{\nu})\right),
\end{equation*}
where
$i$ indicates the four data-taking periods and four dilepton channels,
$\vec{p}$ represents the parameters of interest, 
which are \afour and the various weights (strength factors $\kappa$),
$\vec{\nu}$ is the vector of all nuisance parameters of systematic uncertainties,
$V$ is the covariance matrix of statistical uncertainties, 
which includes the statistical uncertainties in data, signal MC simulations, 
backgrounds (MC simulation and CRs), efficiencies, calibration, and trigger prefiring probabilities,
$D$ is the observed numbers of data events in the $r=(\abs{y},m,\cs)$ bins,
and $T$ is the vector of the corresponding predictions.
For each sample and bin $r$,
\begin{equation*}
T_r(\vec{p},\vec{\nu}) = \sum_{g} S^g_r(\vec{p},\vec{\nu}) + S^o_r(\vec{\kappa},\vec{\nu}) + B_r(\vec{\nu}) ,
\end{equation*}
where $S_r^g$ is the signal contribution from the pre-FSR bin $g=(\abs{Y},M,C)$,
with $C$ being the pre-FSR \cs bin, 
to the reconstructed bin $r$, calculated as
\begin{equation*}
S_r^g(\kappa, \afour, A_0) = S_{r,0}^g \, \kappa \,
\frac{\int_C f(x; \afour, A_0) \, \mathrm{d} x}{\int_C f(x; A_{4,0}, A_{0,0}) \, \mathrm{d} x} ,
\end{equation*}
where $S_{r,0}^{g}$, $A_{4,0}$, and $A_{0,0}$ 
denote reference predictions evaluated from simulation, and
\begin{equation*}
f(x; \afour, A_0) = 1 + x^2 + 0.5 \, A_0 \, (1-3x^2) + \afour \, x .
\end{equation*}
The $S_r^o$ term represents the signal contribution 
from the underflow plus overflow pre-FSR bins to the reconstructed bin $r$.
These predictions are scaled according to floating strength-factor values in 
the underflow and overflow measurement rapidity and mass bins.
The $B_r$ term represents the background contribution.

In each of the nine rapidity bins we have different numbers of mass bins for \afour and $A_0$, 
shown in Table~\ref{tab_bins}, driven by the dilepton mass resolutions. 
For the strength factors, which are more sensitive to bin migrations,
we use wider (and common) mass bins for all rapidity bins.
In addition, the central mass bin of each channel is used for the resolution calibration 
of the reconstructed leptons and, therefore, 
they cannot provide independent strength factor measurements in finer bins. 
Therefore, for the strength factors in the central-central channels, we use 
dilepton mass bins of edges 54, 66, 82, 100, 116, and 150\GeV, 
while one single bin is used for the central-forward channels: 66--116\GeV. 

\begin{table}[h] 
\centering
\topcaption{Numbers of bins and free parameters used in the unfolding.}
\label{tab_parcounts}
\begin{tabular}{ l  c }
\hline
Data-taking periods: & 4 \\
Channels: \uu, \ee, \myeg, \eh & 4 \\
$\abs{y}$ bins in the four channels:  $6+6+4+4$ & 20 \\ 
$\abs{y}$-$m$ bins: $2 \times (6\times11) + 2 \times (4\times9)$ & 204 \\
\cs bins &  20 \\
Bins with more than 10 predicted events & 14\,205 \\[\cmsTabSkip]
\afourym parameters: $3\times (11 + 7 + 3)$ & 63 \\ 
$\kappa(\abs{Y},M)$ parameters: $6\times5  + 4\times1 + 4\times1$ & 38  \\
Nuisance parameters for syst.\ uncertainties & 3\,361 \\
\end{tabular}
\end{table}

Modeling the differential distributions is more challenging than asymmetries. Therefore, we further
decorrelate some of the systematic uncertainties in the unfolding to improve the best-fit $\chi^2$. 
Table~\ref{tab_parcounts} presents the numbers of bins and free parameters involved in the unfolding. 
The main differences with respect to Table~\ref{tab_parameters} are:
(1)~the multijet and \wjet background contributions are decorrelated 
in the measurement $\abs{y}$ and $m$ bins;
(2)~the top quark and \ztt backgrounds are decorrelated in the $\abs{y}$ bins;
(3)~the forward electron efficiencies are decorrelated in $\abs{\eta}$ bins;
(4)~the PDF and $\alpha_S$ uncertainties correspond to 101 eigenvectors of the hessian NNPDF31 PDF set;
(5)~and there are 63 additional nuisance parameters for $A_0(\abs{Y},M)$ 
with a loose prior uncertainty of 50\%, 
which allow us to also fit them in the forward-rapidity bins.
The total number of nuisance parameters is 3361.
In the $\chi^2$ calculation we only include $y$-$m$-\cs bins 
that have a predicted content of more than 10 events.
This set of bins is decided once for the default theory model and is not changed.
The minimization is performed with the L-BFGS method~\cite{lbfgs}, 
which is based on the analytic gradient calculation of the minimization function.
The covariance matrix is evaluated analytically 
by calculating the inverse of the corresponding Hessian matrix.

\begin{figure}[t!]
\centering
\includegraphics[width=\cmsFigWidth]{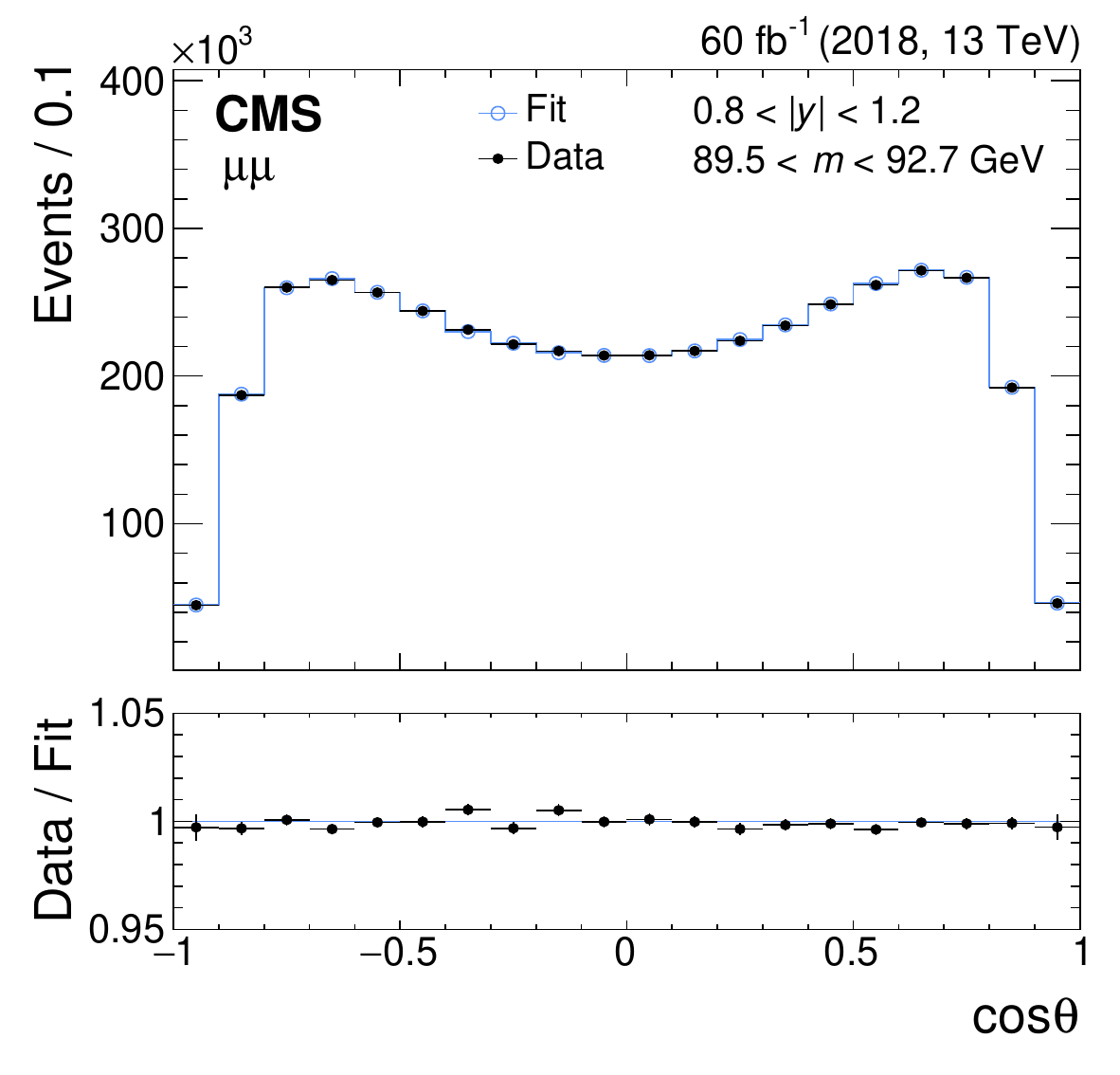}
\includegraphics[width=\cmsFigWidth]{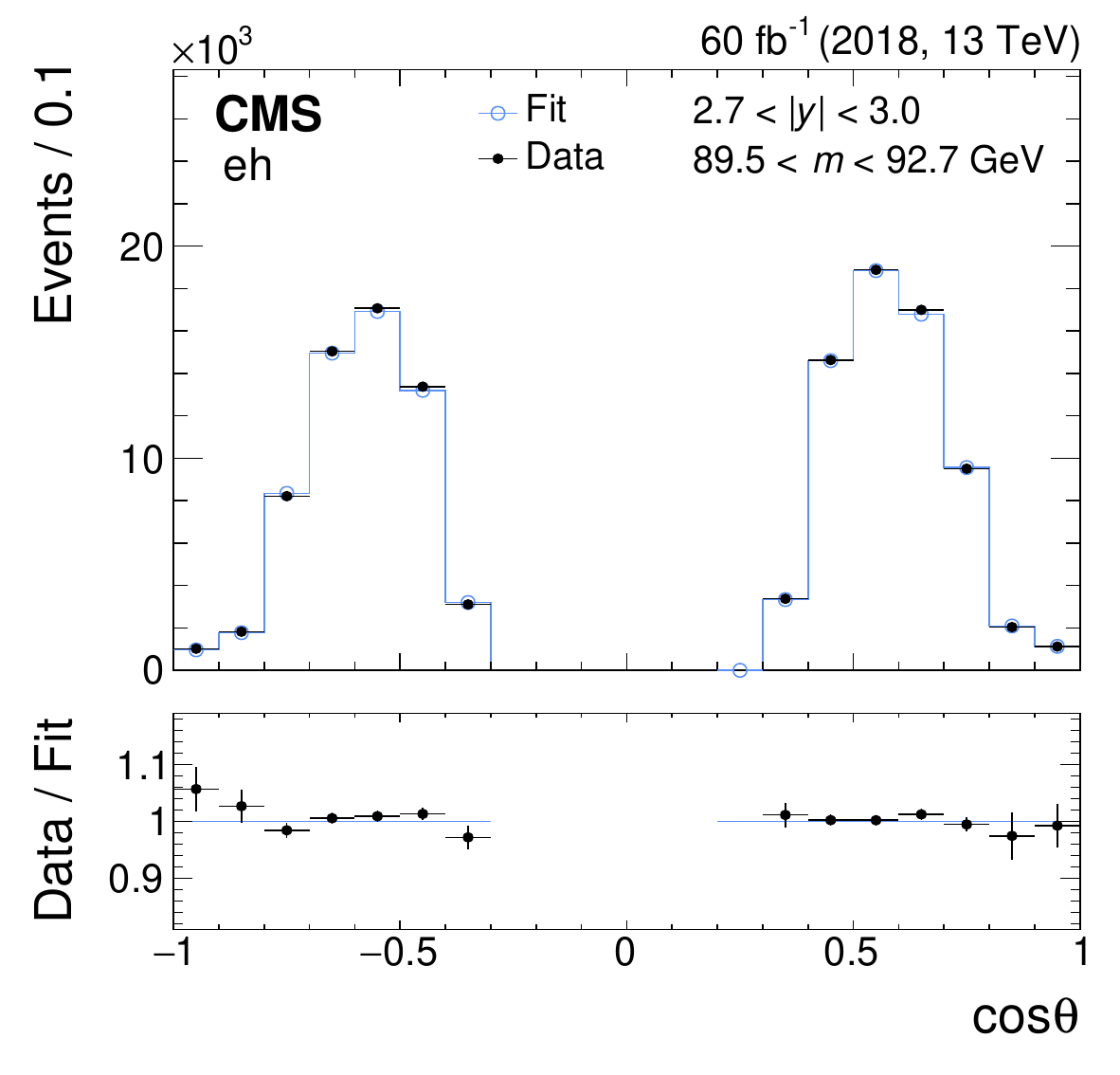}
\caption{Measured and best fit \uu~(\cmsLeft) and \eh~(\cmsRight) 
\cs distributions for 2018,
for the \PZ boson peak and two rapidity bins. 
The error bars represent the statistical uncertainties.}
\label{fig_fitcs}
\end{figure}

Figure~\ref{fig_fitcs} shows the \uu and \eh \cs distributions for the 2018 samples, 
for the \PZ peak and illustrative rapidity bins of the combined fit.
Data in all $y$-$m$ bins, dilepton channels, and years are described very well by the fit.

\begin{figure}[h!]
\centering
\includegraphics[width=\cmsFigWidth]{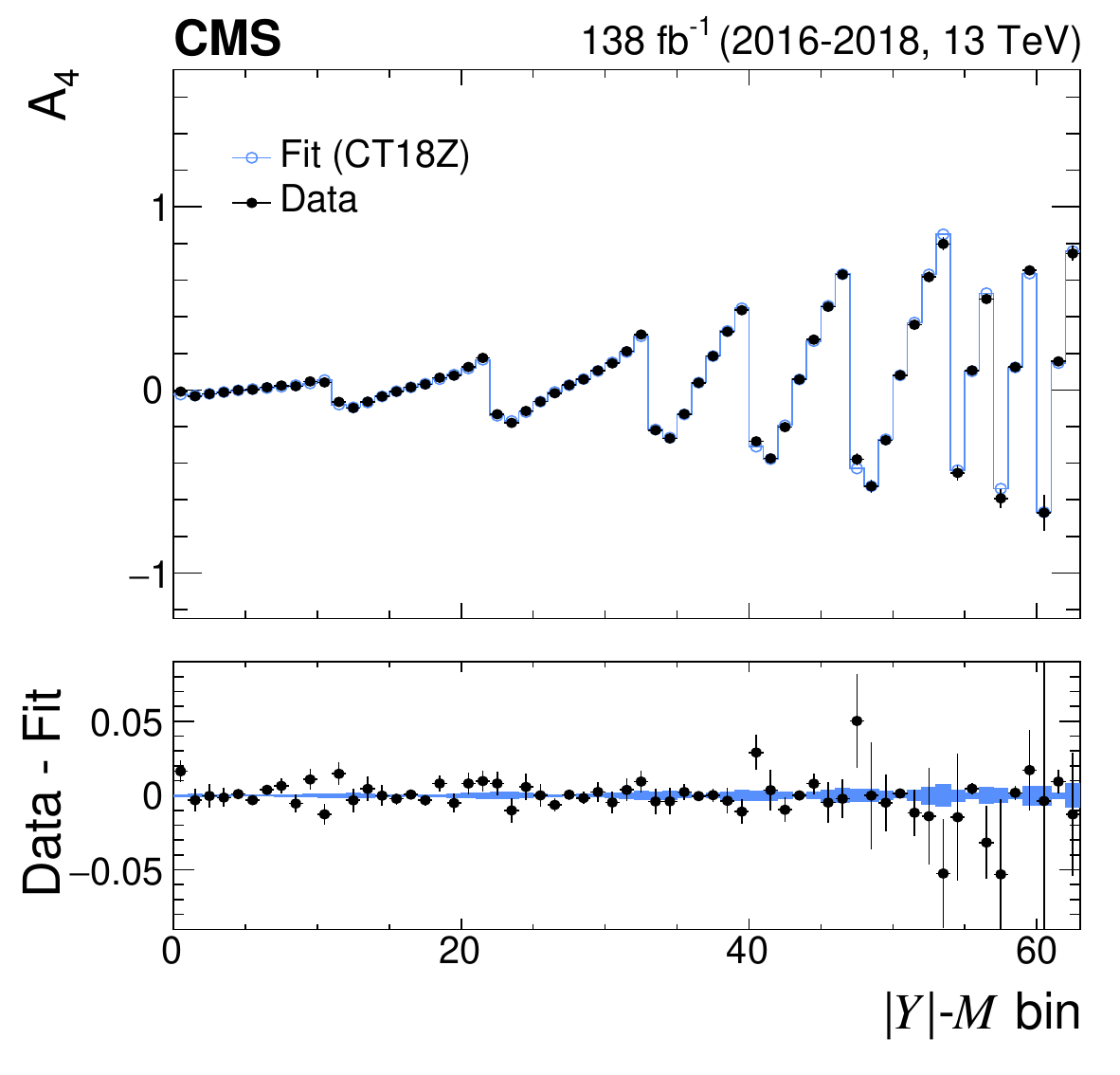}
\caption{Measured and best fit \afourym distributions for the combined 2016--2018 fit with the CT18Z PDF set. 
The shaded band represents the post-fit PDF uncertainty.}
\label{fig_a4}
\end{figure}

We extract \sineff from the measured \afourym using the corresponding templates.
Adding the four channels, we have 63 \afour measurements in $\abs{Y}$-$M$ bins 
(Table~\ref{tab_bins}).
In each fit, we float the \sineff and PDF nuisance parameters.
The small EW and \PZ boson \pt modeling uncertainty 
is included in the covariance matrix calculation. 
For the $\mu_\mathrm{R}$ and $\mu_\mathrm{F}$ uncertainties, 
the fit is repeated for each of the six variations 
and the maximum deviation is used as uncertainty. 
In addition to the default result, obtained with the CT18Z PDFs, 
we also extract \sineff with other recent PDF sets. 
Figure~\ref{fig_a4} shows the measured and best fit \afourym distributions 
for the full data sample.
The corresponding best fit \sineff and $\chi^2$ values are listed in Table~\ref{table_wma}.
The results of the four dilepton channels are compatible with each other.

\begin{table}[ht]
\centering
\topcaption{Measured \sineff values (in units of $10^{-5}$) 
when using the \afourym distributions for the four final-state channels and their sum.}
\label{table_wma}
\begin{tabular}{ c  c  c  c  c }
Channel      & $\chi^2_\text{min}$ & bins & $p$(\%) & \sineff \\ \hline
\uu          & 62 & 54 & 19 & $23\,144 \pm 39$ \\
\ee          & 48 & 54 & 69 & $23\,190 \pm 43$ \\
\myeg        & 11 & 12 & 41 & $23\,249 \pm 60$ \\
\eh          &  9 & 12 & 66 & $23\,123 \pm 48$ \\[\cmsTabSkip]
$\Pell\Pell$ & 64 & 63 & 39 & $23\,154 \pm 32$ \\
\end{tabular}
\end{table}

As a cross-check, we also extracted \sineff by directly fitting the observed \cs distributions. 
The fit configuration is the same as before, except that we do it in a single fit
(instead of the two-step procedure of unfolding and interpretation)
by replacing the free \afourym parameters in the unfolding fit by the values 
calculated as a function of \sineff and PDF parameters.

\section{Results}

\begin{figure}[h]
\centering
\includegraphics[width=\cmsFigWide]{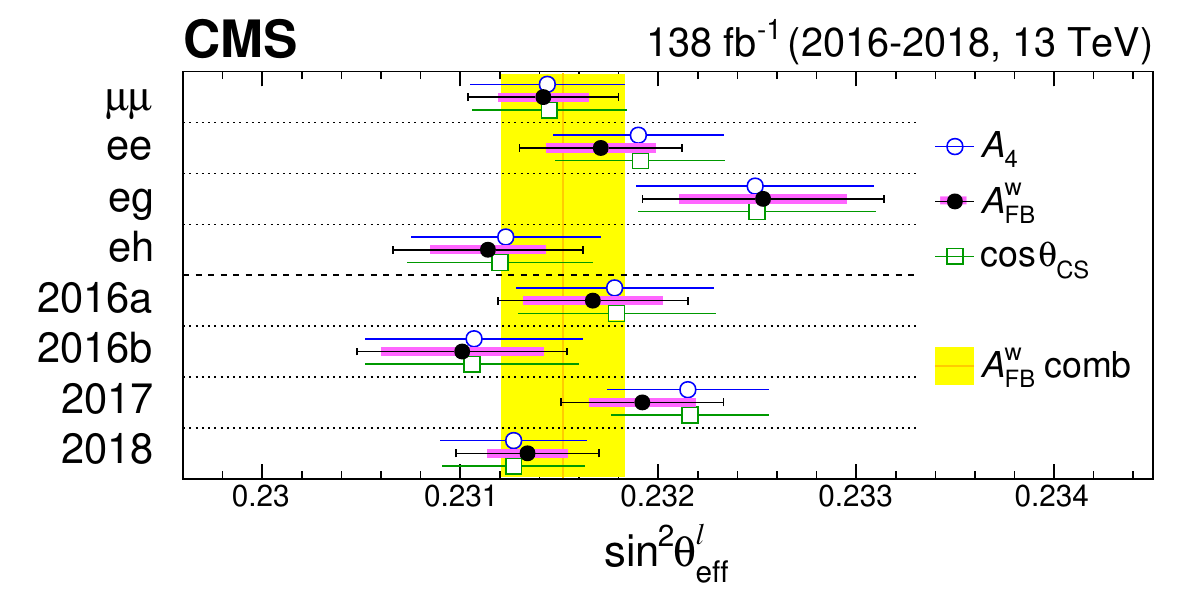}
\caption{Values of \sineff measured with the \afbw, \afour, and \cs fits, 
in each of the four channels using the full 2016--2018 sample (upper)
and in each of the four data-taking periods combining the four channels (lower),
always with the CT18Z PDF set.
The ``comb" band shows the result for all channels and runs combined. 
For the \afbw results, 
the magenta bands show the combined statistical and experimental systematic uncertainties, 
and the black bars represent the total uncertainties.}
\label{fig_res_comb}
\end{figure}

The extracted \sineff values are presented in Fig.~\ref{fig_res_comb},
for the four detection channels (integrating all data)
and separately for each of the four data-taking periods (combining the four channels),
always using the default CT18Z PDF set.
For each case, we show the measurements obtained 
by fitting the detector-level weighted \afbwym and
by fitting the unfolded \afourym angular coefficient of the pre-FSR dilepton;
the result of the direct fitting of the measured \cs distributions is also presented, 
as a cross-check.
The statistical uncertainties in the \afour and \cs results are almost fully correlated;
they are also partially correlated with those of the \afbw result.
Small variations in the results can also be attributed to the different choices of correlations between the 
experimental systematic uncertainties made in the fits, as described in Section~\ref{sec_a4}.
The central values and uncertainties of the three methods are compatible with each other,
given the correlations between samples and systematic uncertainties.

\label{sec:results}

\begin{figure}[t!]
    \centering
    \includegraphics[width=\cmsFigWide]{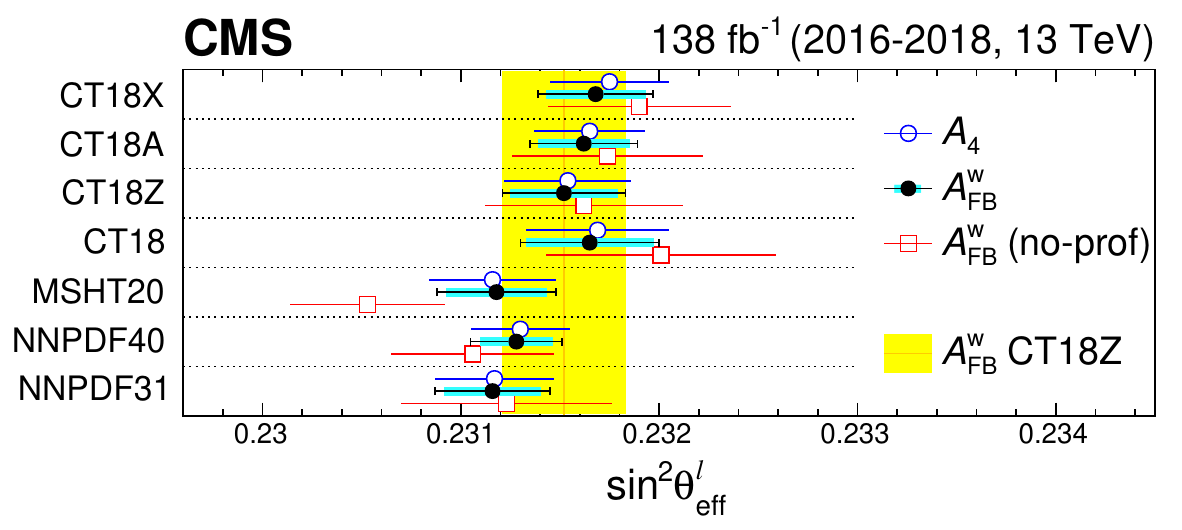}
\caption{Values of \sineff measured with the \afbw and \afour fits,
for seven PDF sets, combining the four channels and using the full 2016--2018 sample.
The orange line and yellow band correspond to the result
obtained with the CT18Z PDFs. 
The red open squares are the results obtained 
without profiling the corresponding PDF uncertainties.
For the \afbw results, 
the cyan bands show the PDF uncertainties
and the black bars represent the total uncertainties.}
\label{fig_pdf_comparison}
\end{figure}

Table~\ref{table_pdf_comparison} and Fig.~\ref{fig_pdf_comparison}
present the \sineff values 
extracted from the measured \afbw or unfolded \afour
when using alternative PDF sets,
combining the four detection channels and the four data-taking periods.
The displayed error bars represent the total uncertainties, 
which include the statistical uncertainty (``stat"), 
and the systematic uncertainties reflecting 
experimental effects (``exp"), the theoretical modeling (``theo"), 
and the PDFs, the latter being the dominating term.
In the case of our baseline result,
corresponding to the \afbw analysis with the CT18Z PDFs,
the individual uncertainties are
$0.00010\stat$, $0.00015\exper$, $0.00008\thy$, and $0.00027\pdf$,
leading to a total uncertainty of 0.00031, 
considering correlations between the various contributions.

\begin{table*}[ht]
\centering
\topcaption{Values of \sineff (in units of $10^{-5}$) 
obtained by fitting the measured \afbw or unfolded \afour,
for seven PDF sets, combining the four channels and using the full 2016--2018 sample.}
\label{table_pdf_comparison}
\begin{tabular}{ l   c c    c c }
PDF & \multicolumn{2}{c}{\afbw (816 bins)}& \multicolumn{2}{c}{\afour (63 bins)} \\ \hline
& $\chi^2_\text{min}$ & \sineff & $\chi^2_\text{min}$ & \sineff\\
NNPDF31\_nnlo\_as\_0118\_hessian~\cite{NNPDF:2017mvq}   & 725 &  $23\,116 \pm 29$ &  61 &  $23\,117 \pm 30$ \\
NNPDF40\_nnlo\_as\_01180\_hessian~\cite{NNPDF:2021njg}  & 731 &  $23\,128 \pm 23$ &  66 &  $23\,130 \pm 25$ \\
MSHT20nnlo\_as118~\cite{Cridge:2023ryv}                 & 736 &  $23\,118 \pm 30$ &  76 &  $23\,116 \pm 32$ \\
CT18NNLO~\cite{Hou:2019efy}                             & 729 &  $23\,165 \pm 35$ &  65 &  $23\,169 \pm 36$ \\
CT18ZNNLO~\cite{Hou:2019efy}                            & 731 &  $23\,152 \pm 31$ &  64 &  $23\,154 \pm 32$ \\
CT18ANNLO~\cite{Hou:2019efy}                            & 731 &  $23\,162 \pm 27$ &  67 &  $23\,165 \pm 28$ \\
CT18XNNLO~\cite{Hou:2019efy}                            & 729 &  $23\,168 \pm 29$ &  64 &  $23\,175 \pm 30$ \\
\end{tabular}
\end{table*}

After PDF profiling~\cite{Bodek:2015ljm}, 
all PDF sets lead to good quality fits and, 
without considering correlations between the PDF uncertainties, 
the results are compatible within uncertainties.
Figure~\ref{fig_pdf_comparison} also shows
the results obtained without profiling the PDFs, 
for which the total PDF uncertainty is evaluated by adding (in quadrature) 
all differences obtained with each PDF error set.
As a result of profiling, we see a significant reduction of the differences 
between the central values of the results obtained with the different PDF sets, 
as well as a decrease of their individual PDF uncertainties.

\begin{figure}[ht]
\centering
\includegraphics[width=\cmsFigWide]{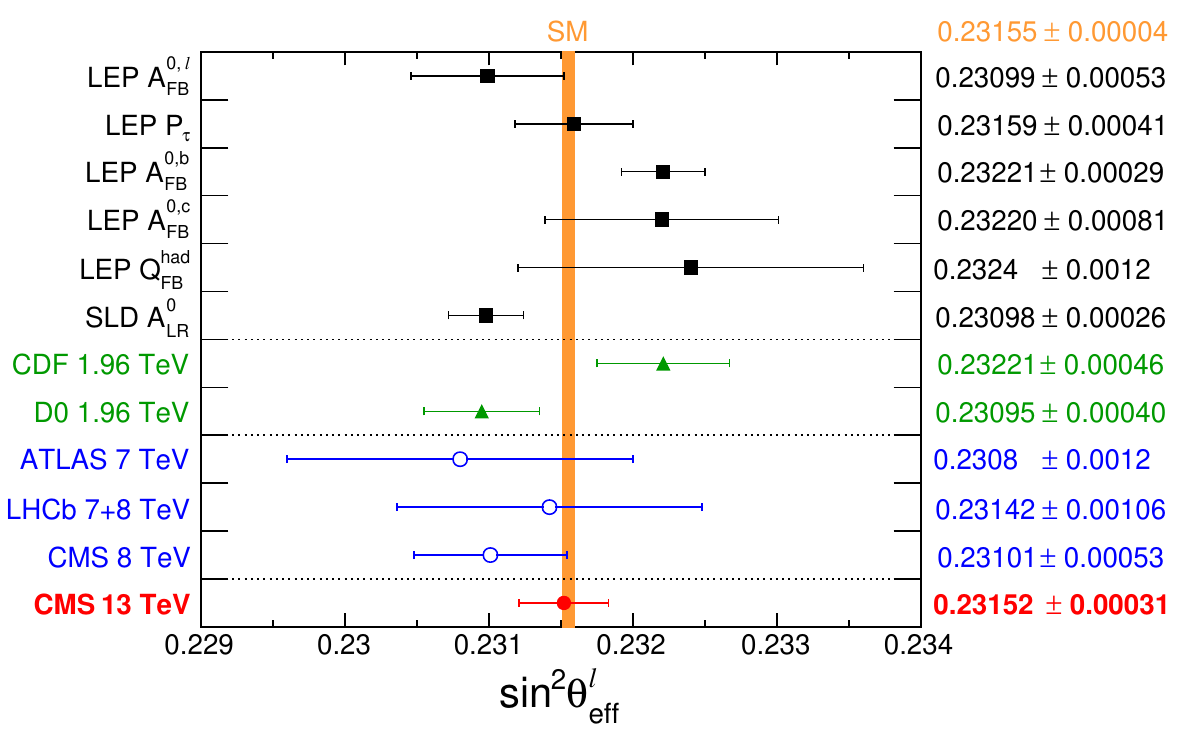}
\caption{Comparison of the \sineff values measured in this analysis
with previous measurements~\cite{lepsld,
Aaltonen:2016nuy,
Abazov:2017gpw,
Aad:2015uau,
Aaij:2015lka,
cms:wma} 
and the result of a SM global fit~\cite{Workman:2022ynf}.}
\label{fig_result}
\end{figure}

Figure~\ref{fig_result} presents the \sineff value 
measured combining the four dilepton channels and using the full 2016--2018 sample. 
This is the most precise result obtained at a hadron collider, 
to date, and agrees with the SM expectation.

In addition, the \sineff based on unfolded \afour is obtained by using \xfitter
open-source code~\cite{Alekhin:2014irh,HERAFitterdevelopersTeam:2015cre}. 
This tool facilitates the reinterpretation of the measurements 
in case the PDF sets or the theoretical models are updated, 
as well as the combination of \sineff values reported by different experiments.

The value of \sineff is determined in a profiling analysis, by minimising the $\chi^2$ function
\begin{equation*}
\ifthenelse{\boolean{cms@external}}{
\begin{split}
& \chi^2(\beta_\text{exp},\beta_\text{th}) = \\
& \sum_{i=1}^{N_\text{data}}
\frac{(\sigma_i^\text{exp}+\sum_j\Gamma_{ij}^\text{exp}\beta_{j,\text{exp}}-\sigma_i^\text{th}-\sum_k\Gamma_{ik}^\text{th}\beta_{k,\text{th}})^2}{\Delta_i^2} \\
& + \sum_j\beta^2_{j,\text{exp}} + \sum_k\beta^2_{k,\text{th}}.
\end{split}
}
{
\chi^2(\beta_\text{exp},\beta_\text{th}) = 
\sum_{i=1}^{N_\text{data}}
\frac{(\sigma_i^\text{exp}+\sum_j\Gamma_{ij}^\text{exp}\beta_{j,\text{exp}}-\sigma_i^\text{th}-\sum_k\Gamma_{ik}^\text{th}\beta_{k,\text{th}})^2}{\Delta_i^2}
+ \sum_j\beta^2_{j,\text{exp}} + \sum_k\beta^2_{k,\text{th}}.
}
\end{equation*}
The correlated experimental (theoretical) uncertainties are included in the nuisance vector $\beta_\text{exp}$ ($\beta_\text{th}$) 
and their impact on the measured distributions (theory predictions) 
is described by the matrix $\Gamma^\text{exp}$ ($\Gamma^\text{th}$).
The index $i$ runs over all $N_\text{data} = 63$ data points of the ($\abs{Y}$-$M$) double-differential \afour measurement,
combining the four detection channels and using the four data-taking periods, 
whereas the $j$ and $k$ indices correspond to the experimental and theoretical uncertainty nuisance parameters, respectively.
The measurements and the uncorrelated experimental uncertainties are represented 
by $\sigma_i^\text{exp}$ and $\Delta_i$, respectively, 
whereas the theoretical predictions are denoted by $\sigma_i^\text{th}$.
The information in the experimental covariance matrix of the double-differential \afour measurement
is included in the $\Gamma^\text{exp}$ matrix.

\ifthenelse{\boolean{cms@external}}{
\begin{table*}[t!] 
\centering
\topcaption{Values of \sineff (in units of $10^{-5}$) extracted by profiling the \afour distribution 
(with 63 data points) using \xfitter, for several PDF sets. 
The reported uncertainties are the total ones, including contributions from the
statistical, experimental systematic, theoretical, and PDF sources.}
\label{xFitter_sineff}
\begin{tabular}{lcc}
    PDF                               & $\chi^2_\text{min}$ & \sineff \\    \hline
    NNPDF31\_nnlo\_as\_0118\_hessian~\cite{NNPDF:2017mvq}   & 61 & $23\,117\pm29$ \\
    NNPDF40\_nnlo\_as\_01180\_hessian~\cite{NNPDF:2021njg}  & 66 & $23\,131\pm24$ \\
    MSHT20nnlo\_as118~\cite{Cridge:2023ryv}                 & 75 & $23\,115\pm30$ \\
    CT18NNLO~\cite{Hou:2019efy}                             & 65 & $23\,166\pm36$ \\
    CT18ZNNLO~\cite{Hou:2019efy}                            & 65 & $23\,152\pm32$ \\
    CT18ANNLO~\cite{Hou:2019efy}                            & 67 & $23\,164\pm28$ \\
    CT18XNNLO~\cite{Hou:2019efy}                            & 64 & $23\,171\pm30$ \\[\cmsTabSkip]
    MSHT20qed\_an3lo~\cite{Cridge:2023ryv}                  & 69 & $23\,137\pm31$ \\
    NNPDF40\_an3lo\_as\_01180~\cite{NNPDF:2024nan}          & 65 & $23\,132\pm25$ \\
    NNPDF40\_an3lo\_as\_01180\_mhou~\cite{NNPDF:2024dpb}    & 65 & $23\,120\pm26$ \\
    NNPDF40\_nnlo\_as\_01180\_mhou~\cite{NNPDF:2024dpb}     & 64 & $23\,110\pm27$ \\
    ABMP16\_5\_nnlo~\cite{Alekhin:2017kpj}                  & 59 & $23\,082\pm25$ \\
    PDF4LHC21\_40~\cite{PDF4LHCWorkingGroup:2022cjn}        & 62 & $23\,133\pm33$ \\
    CT18As\_LatNNLO~\cite{Hou:2022onq}                      & 64 & $23\,108\pm43$ \\
\end{tabular}
\end{table*}

The theoretical uncertainty includes contributions from the missing higher-order QCD and EW corrections, 
as evaluated with \powhegzew, 
and PDF uncertainties, 
evaluated by using grids generated at NLO with 
\MGvATNLO~\cite{Alwall:2014hca} and \textsc{PineAPPL}~\cite{schwan2024nnpdf, Carrazza:2020gss}.
The matrix $\Gamma^\text{th}$ includes nuisance parameters 
reflecting the missing higher-order EW corrections, the PDF Hessian uncertainties, 
and the \sineff parameter itself, which is left free in the fit.
}
{
The theoretical uncertainty includes contributions from the missing higher-order QCD and EW corrections, 
as evaluated with \powhegzew, 
and PDF uncertainties, 
evaluated by using grids generated at NLO with 
\MGvATNLO~\cite{Alwall:2014hca} and \textsc{PineAPPL}~\cite{schwan2024nnpdf, Carrazza:2020gss}.
The matrix $\Gamma^\text{th}$ includes nuisance parameters 
reflecting the missing higher-order EW corrections, the PDF Hessian uncertainties, 
and the \sineff parameter itself, which is left free in the fit.

\begin{table*}[h] 
\centering
\topcaption{Values of \sineff (in units of $10^{-5}$) extracted by profiling the \afour distribution 
(with 63 data points) using \xfitter, for several PDF sets. 
The reported uncertainties are the total ones, including contributions from the
statistical, experimental systematic, theoretical, and PDF sources.}
\label{xFitter_sineff}
\begin{tabular}{lcc}
    PDF                               & $\chi^2_\text{min}$ & \sineff \\    \hline
    NNPDF31\_nnlo\_as\_0118\_hessian~\cite{NNPDF:2017mvq}   & 61 & $23\,117\pm29$ \\
    NNPDF40\_nnlo\_as\_01180\_hessian~\cite{NNPDF:2021njg}  & 66 & $23\,131\pm24$ \\
    MSHT20nnlo\_as118~\cite{Cridge:2023ryv}                 & 75 & $23\,115\pm30$ \\
    CT18NNLO~\cite{Hou:2019efy}                             & 65 & $23\,166\pm36$ \\
    CT18ZNNLO~\cite{Hou:2019efy}                            & 65 & $23\,152\pm32$ \\
    CT18ANNLO~\cite{Hou:2019efy}                            & 67 & $23\,164\pm28$ \\
    CT18XNNLO~\cite{Hou:2019efy}                            & 64 & $23\,171\pm30$ \\[\cmsTabSkip]
    MSHT20qed\_an3lo~\cite{Cridge:2023ryv}                  & 69 & $23\,137\pm31$ \\
    NNPDF40\_an3lo\_as\_01180~\cite{NNPDF:2024nan}          & 65 & $23\,132\pm25$ \\
    NNPDF40\_an3lo\_as\_01180\_mhou~\cite{NNPDF:2024dpb}    & 65 & $23\,120\pm26$ \\
    NNPDF40\_nnlo\_as\_01180\_mhou~\cite{NNPDF:2024dpb}     & 64 & $23\,110\pm27$ \\
    ABMP16\_5\_nnlo~\cite{Alekhin:2017kpj}                  & 59 & $23\,082\pm25$ \\
    PDF4LHC21\_40~\cite{PDF4LHCWorkingGroup:2022cjn}        & 62 & $23\,133\pm33$ \\
    CT18As\_LatNNLO~\cite{Hou:2022onq}                      & 64 & $23\,108\pm43$ \\
\end{tabular}
\end{table*}
}

The \sineff values extracted with \xfitter are listed in Table~\ref{xFitter_sineff}.
The first seven rows report values that can be directly compared to those shown in the 
rightmost columns of Table~\ref{table_pdf_comparison}.
We see that the two sets of values are in agreement,
with small differences in the central values that are attributed to 
the different statistical analysis framework.
The remaining seven rows of Table~\ref{xFitter_sineff} present values obtained with alternative PDF sets:
MSHT20~\cite{Cridge:2023ryv},
NNPDF40~\cite{NNPDF:2024nan,NNPDF:2024dpb},
ABMP16~\cite{Alekhin:2017kpj},
PDF4LHC21~\cite{PDF4LHCWorkingGroup:2022cjn},
and CT18As~\cite{Hou:2022onq}.
The CT18As PDF set assumes asymmetric sea strange quark densities.
The NNPDF40 label ``mhou" indicates that these PDFs sets include ``missing higher-order uncertainties".
The differences in the results obtained with the various PDF sets 
are mostly attributed to differences in the parametrizations and assumptions on the flavor decomposition.

Given the crucial role of the PDFs in the extraction of \sineff 
and the importance of the PDF profiling procedure in reducing the corresponding uncertainty, 
we performed an alternative analysis where the PDFs and \sineff are simultaneously extracted.
Although a global PDF fit based on multiple data sets from different experiments 
is beyond the scope of the present study, a basic QCD fit has been performed, 
considering only the deep inelastic scattering data from HERA~\cite{H1:2015ubc} 
and the present \afour measurement. 
The \xfitter framework is used and the parametrization of the HERAPDF2.0 NNLO 
analysis~\cite{H1:2015ubc} is assumed, as in the profiling exercise. 
The fit is performed at NNLO, applying the \afour theoretical model described above.
The goodness of fit is expressed by a total $\chi^2/\text{ndf}$ of 1457/1194.
The somewhat larger $\chi^2$ of HERA data of 1394/1145, similar to the original 
publication~\cite{H1:2015ubc} is observed, while the CMS A4 measurement contributes with 64/63,
including correlations. 
The  \sineff (both the central value and the uncertainty) is in good agreement~(within 0.00002) 
with that obtained by profiling the HERAPDF20 set, 
giving confidence in the results obtained with the PDF profiling technique.

The results reported in this Letter 
are tabulated in the HEPData record for this analysis~\cite{hepdata}.

\section{Summary}

A precise measurement of the forward-backward asymmetry has been performed,
using proton-proton collisions at $\sqrt{s} = 13$\TeV
collected in 2016--2018 by the CMS experiment and 
corresponding to a total integrated luminosity of 138\fbinv.
The measurement is based on the study of Drell--Yan dimuon and dielectron events.
The effective leptonic electroweak mixing angle \sineff is extracted very precisely 
by fitting the detector-level angular-weighted \afbwym
and the unfolded \afourym angular coefficient of the pre-FSR dilepton,
obtaining compatible results.
Given that the angular-weighted asymmetry method~\cite{Bodek:2010qg} 
benefits from the cancelation of systematic uncertainties in the detection acceptance and efficiencies, 
we use this method for our baseline result.
This measurement has a significantly smaller uncertainty than the previous CMS result~\cite{cms:wma}
because of the larger data sample, an improved analysis technique, 
and the inclusion of central-forward dielectron configurations.
Using the CT18Z set of parton distribution functions we obtain
\ifthenelse{\boolean{cms@external}}{
\begin{equation*}
\begin{split}
\sineff & = 0.23152 \pm 0.00010\stat \pm 0.00015\exper \\
& \pm 0.00008\thy \pm 0.00027\pdf,
\end{split}
\end{equation*}
}{
\begin{equation*}
\sineff = 0.23152 \pm 0.00010\stat \pm 0.00015\exper \pm 0.00008\thy \pm 0.00027\pdf,
\end{equation*}
}
where ``stat", ``exp", ``theo", and ``PDF" denote, respectively,
the statistical uncertainty and the systematic uncertainties reflecting 
experimental effects, the theory modeling, and the PDFs.
Accounting for the correlations between the various contributions,
the total uncertainty, dominated by the PDF term, is 0.00031.
It varies between 0.00024 and 0.00035 depending on the PDF set.
From the unfolded \afourym angular coefficient, 
and using the CT18Z PDF set,
the extracted \sineff value is 
$0.23154 \pm 0.00032$ or $0.23152 \pm 0.00032$, 
depending on the analysis framework.

Our result agrees with the standard model expectation, 
$0.23155 \pm 0.00004$, 
and is the most precise hadron-collider measurement.
The precision is comparable to that of the two most precise measurements
performed in $\Pep\Pem$ collisions at LEP and SLD, 
with respective uncertainties of 0.00029 and 0.00026.
The \afour coefficient, 
measured as a function of the dilepton mass and rapidity, 
can be used in combination with other LHC measurements
or to improve the \sineff measurement using future PDF sets.

\begin{acknowledgments}
    We congratulate our colleagues in the CERN accelerator departments for the excellent performance of the LHC and thank the technical and administrative staffs at CERN and at other CMS institutes for their contributions to the success of the CMS effort. In addition, we gratefully acknowledge the computing centers and personnel of the Worldwide LHC Computing Grid and other centers for delivering so effectively the computing infrastructure essential to our analyses. Finally, we acknowledge the enduring support for the construction and operation of the LHC, the CMS detector, and the supporting computing infrastructure provided by the following funding agencies: SC (Armenia), BMBWF and FWF (Austria); FNRS and FWO (Belgium); CNPq, CAPES, FAPERJ, FAPERGS, and FAPESP (Brazil); MES and BNSF (Bulgaria); CERN; CAS, MoST, and NSFC (China); MINCIENCIAS (Colombia); MSES and CSF (Croatia); RIF (Cyprus); SENESCYT (Ecuador); ERC PRG, RVTT3 and MoER TK202 (Estonia); Academy of Finland, MEC, and HIP (Finland); CEA and CNRS/IN2P3 (France); SRNSF (Georgia); BMBF, DFG, and HGF (Germany); GSRI (Greece); NKFIH (Hungary); DAE and DST (India); IPM (Iran); SFI (Ireland); INFN (Italy); MSIP and NRF (Republic of Korea); MES (Latvia); LMTLT (Lithuania); MOE and UM (Malaysia); BUAP, CINVESTAV, CONACYT, LNS, SEP, and UASLP-FAI (Mexico); MOS (Montenegro); MBIE (New Zealand); PAEC (Pakistan); MES and NSC (Poland); FCT (Portugal); MESTD (Serbia); MCIN/AEI and PCTI (Spain); MOSTR (Sri Lanka); Swiss Funding Agencies (Switzerland); MST (Taipei); MHESI and NSTDA (Thailand); TUBITAK and TENMAK (Turkey); NASU (Ukraine); STFC (United Kingdom); DOE and NSF (USA).

    \hyphenation{Rachada-pisek} Individuals have received support from the Marie-Curie program and the European Research Council and Horizon 2020 Grant, contract Nos.\ 675440, 724704, 752730, 758316, 765710, 824093, 101115353, 101002207, and COST Action CA16108 (European Union); the Leventis Foundation; the Alfred P.\ Sloan Foundation; the Alexander von Humboldt Foundation; the Science Committee, project no. 22rl-037 (Armenia); the Belgian Federal Science Policy Office; the Fonds pour la Formation \`a la Recherche dans l'Industrie et dans l'Agriculture (FRIA-Belgium); the F.R.S.-FNRS and FWO (Belgium) under the ``Excellence of Science -- EOS" -- be.h project n.\ 30820817; the Beijing Municipal Science \& Technology Commission, No. Z191100007219010 and Fundamental Research Funds for the Central Universities (China); the Ministry of Education, Youth and Sports (MEYS) of the Czech Republic; the Shota Rustaveli National Science Foundation, grant FR-22-985 (Georgia); the Deutsche Forschungsgemeinschaft (DFG), among others, under Germany's Excellence Strategy -- EXC 2121 ``Quantum Universe" -- 390833306, and under project number 400140256 - GRK2497; the Hellenic Foundation for Research and Innovation (HFRI), Project Number 2288 (Greece); the Hungarian Academy of Sciences, the New National Excellence Program - \'UNKP, the NKFIH research grants K 131991, K 133046, K 138136, K 143460, K 143477, K 146913, K 146914, K 147048, 2020-2.2.1-ED-2021-00181, and TKP2021-NKTA-64 (Hungary); the Council of Science and Industrial Research, India; ICSC -- National Research Center for High Performance Computing, Big Data and Quantum Computing and FAIR -- Future Artificial Intelligence Research, funded by the NextGenerationEU program (Italy); the Latvian Council of Science; the Ministry of Education and Science, project no. 2022/WK/14, and the National Science Center, contracts Opus 2021/41/B/ST2/01369 and 2021/43/B/ST2/01552 (Poland); the Funda\c{c}\~ao para a Ci\^encia e a Tecnologia, grant CEECIND/01334/2018 (Portugal); the National Priorities Research Program by Qatar National Research Fund;  MCIN/AEI/10.13039/501100011033, ERDF ``a way of making Europe", and the Programa Estatal de Fomento de la Investigaci{\'o}n Cient{\'i}fica y T{\'e}cnica de Excelencia Mar\'{\i}a de Maeztu, grant MDM-2017-0765 and Programa Severo Ochoa del Principado de Asturias (Spain); the Chulalongkorn Academic into Its 2nd Century Project Advancement Project, and the National Science, Research and Innovation Fund via the Program Management Unit for Human Resources \& Institutional Development, Research and Innovation, grant B39G670016 (Thailand); the Kavli Foundation; the Nvidia Corporation; the SuperMicro Corporation; the Welch Foundation, contract C-1845; and the Weston Havens Foundation (USA).
\end{acknowledgments}

\bibliography{auto_generated}
\cleardoublepage \appendix\section{The CMS Collaboration \label{app:collab}}\begin{sloppypar}\hyphenpenalty=5000\widowpenalty=500\clubpenalty=5000\input{SMP-22-010-public-authorlist.tex}\end{sloppypar}
\end{document}

%% file: SMP-22-010-public-authorlist.tex
\cmsinstitute{Yerevan Physics Institute, Yerevan, Armenia}
{\tolerance=6000
A.~Hayrapetyan, A.~Tumasyan\cmsAuthorMark{1}\cmsorcid{0009-0000-0684-6742}
\par}
\cmsinstitute{Institut f\"{u}r Hochenergiephysik, Vienna, Austria}
{\tolerance=6000
W.~Adam\cmsorcid{0000-0001-9099-4341}, J.W.~Andrejkovic, T.~Bergauer\cmsorcid{0000-0002-5786-0293}, S.~Chatterjee\cmsorcid{0000-0003-2660-0349}, K.~Damanakis\cmsorcid{0000-0001-5389-2872}, M.~Dragicevic\cmsorcid{0000-0003-1967-6783}, P.S.~Hussain\cmsorcid{0000-0002-4825-5278}, M.~Jeitler\cmsAuthorMark{2}\cmsorcid{0000-0002-5141-9560}, N.~Krammer\cmsorcid{0000-0002-0548-0985}, A.~Li\cmsorcid{0000-0002-4547-116X}, D.~Liko\cmsorcid{0000-0002-3380-473X}, I.~Mikulec\cmsorcid{0000-0003-0385-2746}, J.~Schieck\cmsAuthorMark{2}\cmsorcid{0000-0002-1058-8093}, R.~Sch\"{o}fbeck\cmsorcid{0000-0002-2332-8784}, D.~Schwarz\cmsorcid{0000-0002-3821-7331}, M.~Sonawane\cmsorcid{0000-0003-0510-7010}, W.~Waltenberger\cmsorcid{0000-0002-6215-7228}, C.-E.~Wulz\cmsAuthorMark{2}\cmsorcid{0000-0001-9226-5812}
% ...
\par}
\vskip\cmsinstskip
\dag:~Deceased\\
$^{1}$Also at Yerevan State University, Yerevan, Armenia\\
$^{2}$Also at TU Wien, Vienna, Austria\\
% ...
